# The IMF Revisited: A Case for Variations

John Scalo

*Astronomy Department, The University of Texas, Austin, Texas 78712*

**Abstract.** A survey of results concerning the IMF derived from star counts is presented, including work up to, but not including, that presented in these proceedings. The situation regarding low-mass stars in the field and in clusters, high-mass stars and intermediate-mass stars in clusters and associations of the Milky Way and LMC, pre-main sequence objects in visible and embedded clusters, and the IMF in galaxies more distant than the Magellanic Clouds is discussed, with an emphasis on the sources of uncertainty. Most of these uncertainties, especially radial mass segregation and unresolved binaries, would steepen the true IMF relative to the apparent IMF. Several cases of apparently large variations in cluster IMFs are pointed out, and a graphical comparison of results for about 60 clusters shows a spread of at least unity in the logarithmic IMF index for all mass ranges above about 1 $M_\odot$. I conclude that either: 1. The uncertainties are so large that very little can be said about an average IMF or IMF variations; or 2. If the observations are taken at face value, there are strong indications of IMF variations, which do not seem to correlate with obvious environmental conditions like metallicity or stellar density. If there is an average IMF, I suggest that it is steepest at intermediate masses. If the variations are real, they offer a useful test of theoretical models.

## 1. Introduction

The organizers have asked me to summarize my impressions of IMF studies since the time of my earlier review paper (Scalo 1986, hereafter S86). The literature in this area has become so large that I must restrict my impressions in various ways. First, I will purposely ignore all the interesting contributions that have been presented during the course of this meeting, restricting the present discussion to the literature that was available prior to the conference. In that way a comparison of the papers included in these proceedings with the questions and problems addressed in this review may serve as a measure of the rate at which progress is being made, and also provide an indication of what specific problems require the most future work.

Second, I restrict this review to studies that directly count stellar (or protostellar or substellar) objects to estimate the IMF. The huge and diverse array of "indirect" IMF constraints based on integrated light properties and chemical evolution arguments are excluded from discussion. It is important, however, to recognize the large number of such constraints. A partial list includes mass-to-light ratios, intensities of gravity-dependent spectral features, correlations between colors, UV fluxes, measures of Lyman continuum flux, ratio of far infrared luminosity to radio continuum flux, near-IR and UV spectral lines that are sensitive to stars in a particular mass range, abundance ratios of elements supposedly produced by different mass ranges, metallicity-gas fraction relation for dwarf irregular and blue compact galaxies, radial abundance gradients in galaxies, abundances in the intracluster medium and their redshift evolution, the tilt of the "fundamental plane" for elliptical galaxies, distributions of properties of field galaxies based on fading burst models for the faint blue galaxy population, the color distribution of galaxies at intermediate and large redshifts, and others. Since star formation history, IMF, metallicity, stellar interior and atmosphere model uncertainties, and galaxy evolution model assumptions are all folded into most of these constraints, it should be clear that simultaneous use of multiple constraints is required. Even then, an examination of the large literature on evolutionary synthesis, which does use multiple constraints, illustrates the large uncertainty to be expected. I have also omitted from discussion attempts to model the IMF theoretically; that will be the subject of a separate review.

Even given these restrictions, the size of the literature required that I be somewhat selective, and I am sure there are many significant papers that I have overlooked. In particular, I have concentrated on recent work, often without reference to earlier studies that form the context of current research. Much of this work prior to 1985 was discussed in S86. I have also omitted discussion of the luminosity functions and IMFs of stars in globular clusters and halo stars in the solar neighborhood, and results of gravitational microlensing experiments. Hopefully these subjects will be discussed elsewhere in this volume.

My general impressions of the field can be summarized simply. First, there have been tremendous advances in methodology in the past decade, as outlined below. These advances include more sophisticated instrumentation (e.g. HST, IR cameras) and analysis techniques (e.g. treatments of incompleteness and contamination in studies of star clusters), a recognition of the need for certain types of data (e.g. spectra for high-mass stars), more detailed studies of stellar evolution and atmospheres, and examination of various physical effects (e.g. inflections in the mass-luminosity relation, the effect of unresolved binaries). The net effect of all these advances has been, at least from my view, an accentuation of the large uncertainties present in all IMF estimates.

My second general impression is that, contrary to some views currently held, if the existing empirical estimates of the IMF are taken at face value, they present strong evidence for IMF variations, and these variations do not seem to depend systematically on physical variables such as metallicity or stellar density. I see no evidence that a meaningful average IMF index can be inferred for any mass range, except possibly for low-mass stars. An alternative interpretation is that systematic uncertainties in the empirical results are so large that the form of the IMF may be universal, but is essentially unknown.

In what follows I refer to the "initial mass function" as the differential probability distribution of stellar mass m per unit *logarithmic* mass interval, denoted F(log m). Because we are only interested in the shape of this function, the normalization is arbitrary, except in the case of field stars, where the facts that stars of different masses can be observed to different limiting distances and the dependence of galactic scale height on age (and hence mass) requires a normalization per unit area of the disk. I use the term "initial mass spectrum" f(m) to refer to the differential distribution of masses per unit mass interval, proportional to F(log m)/m. I use only F(log m) in this paper. It is convenient and useful to characterize the IMF in a particular range of masses by the local logarithmic derivative $\Gamma \equiv$ dlogF(logm)/dlogm, which I call the local "index" of the IMF. The danger of this characterization is that it encourages the assumption that the IMF is a power law with constant $\Gamma$, when the data often suggest a double or multiple power laws or even ledges, dips, or peaks; several examples are given below. In many cases the data are such that a single power law seems like the simplest fit that is warranted. However I urge that researchers attempt to apply some statistical test of the validity of a single power law fit, like the Kolmogorov-Smirnov test used by Petitjean et al. (1993) in a different context, where clear evidence for a double power law was found. (See also Kroupa et al. 1991, 1993.) In what follows I summarize various results as if a single power law with index $\Gamma$ characterizes the data, but it should be noted that in very few cases have such a statistical test been carried out.

Another point worth noting concerns the common practice of estimating the IMF using a least squares fit to a histogram of the number of stars in each log mass bin. Other estimators of the IMF, such as maximum likelihood methods (e.g. Willmer 1997) and non-parametric kernel estimators (e.g. Vio et al. 1994), which do not require histogram binning, are available, and are generally believed to yield more robust estimates of probability distributions, although they have their own problems, such as the noise amplification pointed out by Gould et al. (1997) for a maximum likelihood method. An important and much-overlooked paper by Tarrab (1982) derived maximum likelihood IMF indices by counting stars between evolutionary tracks for 75 open clusters. It is presently unknown whether the application of such methods to the data reviewed below would significantly alter the results. The few recent IMF papers in which alternative estimation techniques are used are mentioned below.

The plan of this review is to first discuss the severe problems involved in the estimation of the field star IMF in sec. 2, which concludes that, with some exceptions, clusters and associations provide a more suitable site for the empirical study of the IMF. Sec. 3 summarizes recent estimates of the IMF for massive ($\gtrsim 10 M_\odot$) stars in clusters and associations using spectroscopy and evolutionary tracks. The main point, which is the overall theme of this review, is that it is difficult to decide between an interpretation that sees the data as consistent with a universal high-mass IMF within the large uncertainties and one that sees significant IMF variations. This conundrum becomes more apparent in the discussion in sec. 4 of the intermediate-mass (1–15 $M_\odot$) IMFs derived for a number of clusters and associations in the Milky Way and the LMC, where evidence for significant variations must be weighed against equally compelling evidence for uncertainties due to radial mass segregation, completeness and contamination corrections, and unresolved binaries. In sec. 5 techniques and problems involved in IMF estimation for clusters of (mostly) pre-main sequence (PMS) objects are reviewed. A brief discussion of our scanty information on the IMF in galaxies more distant than the Magellanic Clouds, based on star counts, is given in sec. 6. A graphical summary of all the results discussed for clusters and associations is given in sec. 7.

## 2. The Field Star IMF

Sec. 2 of S86 gave a detailed derivation of the field star IMF, using the Salpeter method. Rather than rely on Miller & Scalo (1979, MS79), I attempted to estimate the required ingredients, as much as possible, "from scratch." These ingredients include the luminosity function (LF), the mass-luminosity (m-L) relation, main sequence lifetimes, the relation between scale height and mass, the correction for evolved stars, and the history of the galactic star formation rate (SFR). In particular, 15 different LFs covering different mass ranges were examined (not counting the additional 5 LFs for massive stars compared in sec. 2.6.4.). The two most important conclusions reached were the following. 1. The field star IMF does not exhibit the lognormal form found in MS79. For masses $m \gtrsim 1$ the IMF was roughly a power law, and even showed signs of flattening with increasing mass, rather than steepening as in MS79. It has been a source of consternation for me that many authors, especially in the areas of PMS stars and galaxy evolution, continue to use the lognormal IMF advocated by MS79. There is no longer any evidence I know of for a lognormal IMF. 2. The uncertainties in the field star IMF are huge. It was for this reason that I purposely did not present some general-purpose functional fit to the derived IMF. Instead I proceeded to examine the IMF derived from clusters and associations (sec. 3), star counts in nearby galaxies (sec. 4), and constraints from the integrated light of stellar populations (sec. 5) and chemical evolution (sec. 5). The uncertainties associated with all these approaches, as in the case of the field star IMF, were so large that they could all accommodate a universal IMF, except possibly for starburst regions. Worded more appropriately, there was insufficient evidence to establish IMF variations. However it should be noted that in the discussion of clusters I concluded that $\Gamma$ for 1–10 $M_\odot$ is –1.5 to –1.8, but probably flatter at larger masses, suggesting $\Gamma = -1.2$ there (see summary on p. 157 and Fig. 30 on p. 131).

With some qualification, there has apparently been no updated treatment of the field star IMF covering the entire range of stellar masses. Kroupa and collaborators (Kroupa et al. 1990, 1991, 1993, Kroupa 1995a-d, Kroupa & Tout 1997; see Kroupa 1997 for a review) and others (see below) have examined in detail how the inflections in the m-L relation at $m \approx 0.7$ (importance of $H^-$ opacity source) and $m \approx 0.3$ ($H_2$ formation) could produce features in the LF which are not reflections of features in the IMF, and how the presence of unresolved binaries and Malmquist bias could account for the differences between the LF estimated from photometric samples and the LF based on nearby stars with trigonometric parallaxes. But this work was entirely concerned with stars less massive than 1 $M_\odot$; for higher mass stars the IMF was taken directly from S86. Several other recent papers have estimated the very low-mass field star IMF (e.g. Tinney et al. 1992, Bessel & Stringfellow 1993, Jarrett et al. 1994, Mera et al. 1996, Gould et al. 1997); some of these are discussed below.

Basu & Rana (1992) did re-examine a number of the ingredients that go into the field star IMF. However it is difficult to evaluate the validity of their result because of the LFs employed. Although purporting to use LF determinations more recent than used in S86, tracing the sources back through Rana (1987) shows that, for $m \gtrsim 0.8$ the LFs are mainly from older work. For example, the Gilmore & Reid (1983) LF they use in this range is actually taken from MS79;

the Eaton et al. (1984) LF quoted is actually from Allen's (1973) *Astrophysical Quantities*, itself a summary of much older work (not a transformation of the Eaton et al. near-IR LF); the Robin & Creze (1986) LF they quote is also from Allen (1973) for $M_V < -2$ ($m > 8$). For $M_V > 6$ ($m \lesssim 0.8$) Basu & Rana use Stobie et al.'s (1989) LF, but for larger masses the LFs are of questionable quality, and even for small masses their adopted LF has been largely superseded. Basu & Rana do give a useful discussion of scale heights and the m-L relation. A recent detailed discussion of the ingredients needed for the field star IMF is given by Haywood et al. (1997a,b), although they were mostly concerned with the $m \approx 1-3$ mass range, where the uncertainty in SFR history plays a major role.

Viewed in the light of subsequent work, there were (at least) two major problems with the field star IMF derived in S86. The first concerns low-mass stars. The derivative of the m-L relation, required for conversion from LF to IMF, shows significant structure at $m \approx 0.7$ and m$\approx 0.3$, which led to the erroneous conclusion that there was a minimum and a maximum in the IMF at these masses, respectively. These inflections in the m-L relation were first pointed out by D'Antona & Mazzitelli (1983), emphasized by Kroupa et al. (1990), and recently studied in more detail, including metallicity-dependence, by D'Antona & Mazzitelli (1996), Alexander et al. (1997), and Chabrier et al. (1996). The latter paper showed the importance of using non-grey model atmospheres as boundary conditions for the stellar evolution models. Kroupa & Tout (1997) have discussed the current situation in detail. A recent paper by Malkov et al. (1997) emphasizes that there may still be considerable uncertainty in the m-L relation.

There is now considerable evidence that the turnover observed in the field star (and cluster) LF can *at least partially* be accounted for by the decrease in slope of the m-L relation at $m \approx 0.3$ (see Kroupa 1995a,b), especially considering the finding of von Hippel et al. (1996) that the LF peak in clusters is correlated with metallicity in the way expected for a single star of $m \approx 0.3$, a result confirmed in more detail by Kroupa & Tout (1997). Following von Hippel et al., a turnover in the IMF at low masses is still possible, but may not be necessary. This conclusion depends on the LF adopted for nearby stars, as shown in Reid & Gizis (1997), who do find evidence for a turnover at $m \sim 0.1$ (see below).

It still appears that, despite these advances, we are far from understanding the form of the field star IMF at very low masses around the hydrogen-burning limit. Mera et al. (1996) derived the field star low-mass IMF using the m-L relation of Chabrier et al. (1996), based on models by Baraffe et al. (1995), and the local parallax field star LF of Kroupa (1995a,b), and found a steep increase with decreasing mass beginning around 0.15 $M_\odot$, for both the parallax and photometric LF samples. They suggest $\Gamma \approx -1$ for the parallax sample, which should not be depressed by the effects of unresolved binaries. Inspection of Chabrier et al.'s Fig. 4 shows that this steep increase in the IMF is a consequence of the significant decrease in the derivative of the theoretical m-L relation that occurs below about 0.15 $M_\odot$. However this relation assumes an age of 5 Gyr (and solar metallicity) so that all objects with $m > 0.07$ are already on the main sequence. But a significant fraction of stars in the solar neighborhood are younger than 5 Gyr (see, e.g. Noh & Scalo 1990), and these PMS stars will lie at larger luminosities for a given mass. The contribution of these stars will reduce the flattening of the m-L relation. This dependence of the derivative of the m-L relation on adopted age can be seen clearly in Fig. 3a of Tinney et al. (1992). Chabrier et al. point out that the three lowest-mass objects in the empirical m-L relation show signs of youth and have larger luminosities than predicted by their theoretical m-L relation, confirming the direction of the effect. It therefore seems to me that a valid IMF determination for the very low-mass ($\lesssim 0.1\ M_\odot$) field stars will require counting between PMS evolutionary tracks in the H-R diagram. Unfortunately the number of stars in the parallax sample is so small that we cannot tell whether they properly represent the average distribution of ages. If the three lowest-mass bins ($\lesssim 0.1\ M_\odot$) of Mera et al. are omitted for these reasons, the remaining IMF points are well-fit with $\Gamma \approx -0.3$, the value advocated by Kroupa et al. (1993) for $m = 0.1$ to $0.5$.

The problem with the very low-mass field star IMF involves, besides the m-L relation, the estimation of the local LF, as shown by the recent paper by Reid and Gizis (1997, RG). The nearby-star parallax sample used by Mera et al. (1996), taken from Kroupa (1995a,b) contains more stars at faint magnitudes $M_V \gtrsim 14$, than the 8-parsec sample used by RG, as shown in RG Fig. 4. Unfortunately the 8pc sample is a mixture of trigonometric and spectroscopic parallaxes, with the latter based on a TiO-$M_V$ correlation that exhibits large scatter, introducing potentially large distance uncertainties. This 8-parsec sample of RG has an IMF that is flat ($\Gamma \approx 0$) between 0.1 and 1 $M_\odot$, with evidence for a decline ($\Gamma > 0$) below 0.1 $M_\odot$, and no evidence for the steepening ($\Gamma = -1.2$) found in the 0.5–0.9 $M_\odot$ mass range by Kroupa et al. (1990, 1991, 1993). This is for a fit to the Henry and McCarthy (1993) empirical m-L relation that is apparently biased toward young and probably overluminous stars at the lowest masses. As noted above, the derived IMF at lowest masses will depend on assumptions about the age distribution. Another problem is that the adopted fit contains discontinuities, and it is unclear how the derivative of this fit (which is what enters the IMF estimate) was treated. RG compare with several theoretical m-L calibrations, and show that the Baraffe and Chabrier (1995) models, used by Mera et al. (1996), and which should be the most physically realistic because of the use of non-grey atmospheres, are discrepant in the sense that the derived $\Gamma$ depends on whether $M_V$ or $M_{bol}$ is used in the m-L relation; Chabrier et al (1996) had noted that the $M_K$ results were discrepant. Mera et al. (1996) had used $M_V$, which gives the steepest $\Gamma (= -0.5$ in RG), but $M_K$ seems more appropriate since it is where most of the stellar flux is emitted. Using the Burrows et al. (1993) m-L relation for $m < 0.2$, RG find $\Gamma = -0.3$. Malkov et al. (1997), in a comparison of different sets of models with the observed m-L data, found that the Baraffee et al. models disagreed with observations for very low masses, even when $M_V$ was used. However Kroupa & Tout (1997) present other evidence that favors the Baraffe et al. models, such as the effective temperatures.

The debate over the question of why the photometric survey LF differs from the local parallax LF is complicated. RG find that the mass ratio distribution, which is a crucial ingredient for estimating the magnitude of the effect of unresolved binaries, is much different for M dwarfs than for the G dwarf mass ratio distribution used by Kroupa (1995a,b), and conclude that unresolved binaries cannot acount completely for the discrepancy. Instead they suggest that it is uncertainties in the color-magnitude calibrations of the photometric parallax analyses that are responsible. However it is not so clear that there is disagreement between the two mass ratio distributions (Kroupa, private communication). In any case, all parties seem to now agree that the local parallax sample provides the most reliable estimate of the IMF at small masses, and most agree that $\Gamma$ is between 0 and –0.5 for 0.1 to 1 $M_\odot$. The question of a turnover or rise at $\sim 0.1 M_\odot$ remains in dispute, but it seems to me that this can never be resolved from the parallax sample because the m-L relation depends sensitively on age at low masses, and even if we know the ages of the stars in the sample and used evolutionary tracks to derive the masses instead of the m-L relation, the number of very low-mass stars in the current parallax sample is simply too small for that sample to be regarded as statistically representative of the IMF, not only in terms of counting uncertainties and spatial fluctuations, but as representatives of the age distribution. Thus the situation for the low-mass field star IMF is that there is some convergence in the sense that most recent estimates give $\Gamma$ between 0 and –0.5 for 0.1 to 1 $M_\odot$, if we omit the 3 lowest mass bins of Mera et al. (1996) for reasons mentioned above. However the dependence of the lowest mass IMF on star formation history because of the presence of PMS stars seems to preclude a determination of whether or not a turnover occurs.

HST photometric LF estimates could significantly improve the statistical situation at very low masses, but the number of stars is still small, the unresolved binary correction, while possibly reduced somewhat, is still important, and besides, the derived LF is apparently uncertain because of the calibration of the photometric data ( Santiago et al. 1996). Perhaps a combination of HST photometry with the ingenious "opaque screen" method of Jarrett et al. (1994), which avoids several difficulties, including Malmquist bias, would be useful.

A recent discussion of HST results accumulated to date is given by Gould et al. (1997), who find a rather clear change of slope of the IMF at about 0.6 $M_\odot$. Above this mass the IMF, based on the local LF of Wielen et al. (1983), has an index $\Gamma = -1.2$ up to $m = 1.2$, but the derived IMF decreases with decreasing mass ($\Gamma$ positive) below 0.6 $M_\odot$. The value of the index between 0.6 and 1.2 $M_\odot$ should be given little weight: Not only is the LF uncertain in that range (see S86, Fig. 4), but the derived IMF depends sensitively on the assumed galactic age and SFR history (S86, Fig. 17, see below). Gould et al. apparently assumed no birthrate correction was necessary, which makes the derived index relatively meaningless. For the low-mass stars Gould et al. recognize the seriousness of the effect of unresolved binaries, and a crude correction for the effect gives $\Gamma \approx +0.1$ for 0.1 to 0.6 $M_\odot$. The "break" in $\Gamma$ at $m \approx 0.6$ may be real, and roughly agrees with some other studies, in particular with the Pleiades IMF derived by Meusinger et al. (1996; see Fig. 2 below) and the field star IMF of Kroupa et al. (1991, 1993) which was based on a parallax sample. It will be suggested below that the IMF in the 1–3 $M_\odot$ range exhibits large cluster-to-cluster variations, so all field star IMF results in this range would have to be interpreted as some ill-defined mean. However the reality of the "break" is questionable because it is so near the mass above which corrections for finite stellar lifetimes ($\sim 0.85$ $M_\odot$) become crucial and extremely uncertain (S86, Fig. 17), an effect apparently neglected by Gould et al., and because the break occurs at a mass where their adopted m-L relation is discontinuous (Kroupa, private communication). It is not clear what to make of the rise in the Gould et al. IMF below 0.1 $M_\odot$, which is similar to that shown in Mera et al. (1996). The uncertainty involved in using a (poorly-defined) m-L relation to represent a population that contains a significant fraction of PMS stars with an uncertain age distribution remains a possible explanation, and a major problem.

It seems to me that at present the problems for low-mass field stars are so severe that the IMF around 0.1 $M_\odot$ and below must come from open clusters, where the star formation history is not as severe a problem, although in that case radial segregation, unresolved binaries, and other effects are important, as discussed below.

Some of these uncertainties could be avoided by concentrating on very old clusters, for which all stars have the same age and should be on the main sequence. Globular custers provide a useful test, although they have their own specific problems. Without attempting to review the globular cluster literature, it is worth noting that the recent study of two globular clusters by Di Marchi & Paresca (1997) concludes that the IMF of these clusters flattens out and probably drops below 0.2 $M_\odot$. However globular clusters are very dynamically evolved, and the relation between present-day mass functions and initial mass functions is far from clear (see Vesperini & Heggie 1997).

A point of particular interest, and confusion, has been the appearance of a peak (for monotonically decreasing SFR) or plateau (for constant or increasing SFRs) at $m \approx 1.2$ in S86 (Fig. 17). The discussion of the reality of this feature was based on the reality of the apparent minimum of the IMF at $m \approx 0.7$, which was due to a minimum in the LF at $M_v \approx 7$. The possibility that this minimum in the IMF might be an artifact due to a neglect of a flattening of the m-L relation, as originally suggested by D'Antona & Mazzitelli (1983), was rejected in S86, a rejection that now clearly appears in error, as emphasized by Kroupa et al. (1990) and others. However even when this minimum is removed, the peak or plateau at $m \approx 1.2$ remains, for any monotonically varying SFR: It is not an artifact of the m-L relation, as claimed by Kroupa et al. (1993) and others. In S86 I suggested that this feature may be evidence for some sort of bimodal star formation process. I later realized (Scalo 1987, 1988) that this peak or plateau could be the signature of a peak in the galactic SFR around 6 Gyr ago, a peak that was found independently from the chromospheric age distribution (Barry 1988). This SFR peak can be seen, at lower age resolution, in Twarog's (1980) unsmoothed data (see Noh & Scalo 1990) and has been inferred from the galactic metallicity distribution (see Rocha-Pinto & Maciel 1997). The field star IMF around $m \approx 1$ is hopelessly entangled with the galactic SFR history. This can be seen in the recent attempt to derive constraints on the IMF in the 1–3 $M_\odot$ range from color distributions of field stars by Haywood et al. (1997a,b). The dependence of the derived IMF on SFR history precludes any estimates of $\Gamma$ for field stars in the mass range 0.85 to 1.3 $M_\odot$. The IMF in this mass range can only be estimated from open cluster IMFs. Since the open cluster IMFs show no evidence for the field star feature around $m = 1.2$ (see sec. 4 below), I conclude that the feature in the field star IMF is an artifact due to a peak in the galactic SFR around 6–8 Gyr ago.

A second area of major development concerns the IMF of massive stars. Massey, Parker, Garmany, and coworkers have shown convincingly that the LF is inadequate for estimating the IMF of massive stars, whether in the field or in clusters and associations. First, because the bolometric correction is a strong function of effective temperature for massive stars, and because of their significant evolution to cooler temperatures, stars in a given absolute visual magnitude range will represent a mixture of masses; there is no one-to-one correspondence between $M_V$ and mass. Second, optical and even near-UV colors only sample the low-frequency tail of the energy distributions of these higher temperature stars, and so colors are insensitive to effective temperature. This means that even a comparison of an observed color-magnitude diagram with theoretical evolutionary tracks is incapable of yielding a reliable IMF. Spectroscopy is necessary.

The inadequacy of the LF for IMF estimates of massive stars has been convincingly demonstrated by Massey et al. (1995), and was first emphasized by Massey (1985). A major achievement in IMF studies has been the decade-long effort by Massey and collaborators to obtain spectra of large numbers of OB stars in our Galaxy and nearby galaxies, and to compare the resulting H-R diagram with evolutionary tracks. This approach is not without its own problems (e.g. small sample size, uncertainties in stellar evolutionary tracks), as discussed below. Still, the spectroscopic approach is the only viable method for direct estimates of the IMF of massive stars, except possibly for integrated light constraints.

In S86 the IMF for massive stars was determined from the LF of stars in the sample of Garmany et al. (1982) and compared with the IMF derived by Garmany et al. using the spectroscopic approach, mostly as an exercise to find the degree to which the two methods agreed. The agreement that was found between the two methods must be considered fortuitous. In a separate section of S86 devoted to massive stars, a number of IMFs were derived from different high-mass star samples, and a huge range in the IMF slope was found. Of particular interest is the result from the study of Van Buren (1985), since he used a sample that was segregated by spectral type and luminosity class, approximately the spectroscopic approach, and because Van Buren employed a reddening model that was much more detailed than other work. The result was an IMF slope of –1.1 to –1.4 for massive stars (depending on the importance of convective overshoot on the m-L relation; see S86 sec. 2.3 and 2.4), in better agreement with the more recent massive star IMFs estimated by direct spectroscopic matching with evolutionary tracks (see below). In any case, the IMF slope for massive stars derived from the m-L relation of the Garmany et al. sample given in S86 is now extinct, just as is the S86 IMF for masses below 1 $M_\odot$.

The upshot of all this is that the field star IMF in S86 is only valid, if at all, in the mass range 1.5 $M_\odot$ to 15–20 $M_\odot$. It is worth noting that the S86 IMF in this mass range is "convex," rather than "concave" as in MS79. For 1.4–3.5 $M_\odot$, $\Gamma \approx -2.6$, and for (1.9 or 2.5)–12 $M_\odot$, $\Gamma \approx -1.8$. These results are relatively independent of the assumed star formation rate history. However even in that mass range the uncertainties are large, and it seems much more desirable to use individual clusters and associations to determine the IMF.

## 3. Massive Stars in Clusters and Associations

As discussed above, it is now clear that the IMF for massive stars cannot be reliably estimated from photometry alone: spectroscopy is required. The insensitivity of the LF to changes in the IMF is still one of degree, however. As pointed out by Bresolin et al. (1996) in their study of M 101, the similar LF slopes found for the brightest stars in a number of galaxies (see references given in Bresolin et al.; also Malumuth et al. 1996) can probably be used to rule out any *extreme* variation in the high-mass IMF slope, although the actual degree of variation allowed has not been examined. The arduous task of acquiring spectra for a significant sample of OB stars in clusters and associations in the Milky Way and LMC has been carried out mostly by the Boulder group. Massey, Johnson, & DeGioia-Eastwood (1995) presented new data for 10 OB clusters and associations in the Milky Way and updated earlier work on three more such groups. Massey et al. also summarized work on five LMC regions. A recent review is given in Massey (1997). There are still uncertainties in the results due to, for example, the evolutionary tracks, that are difficult to estimate (see Maeder & Conti 1994 for a concise review), but the largest uncertainty lies in the relatively small sample size, which results in only a small number of usable mass bins and relatively large Poisson errors for the highest-mass bins. I discuss first the results for the Milky Way groups.

The number of stars used in the IMF estimates for these clusters varies from 6 to 93. Derived values of $\Gamma$ are very uncertain since they are based on fits to only 3 to 5 (in one case) mass bins, and often the largest mass bin contains only 1 or 2 stars. The quoted uncertainties in $\Gamma$ range from 0.2 to 0.6. The values of $\Gamma$ range from –0.7 to –2.1, with a weighted mean of –1.1. Massey et al. were "more struck by the consistency in the values rather than the range. The clusters with the most extreme values also show the largest formal $1\sigma$ errors on the slopes." This latter statement is apparently not true, since the two associations with $\Gamma = -0.7$ and –0.8 have quoted uncertainties of only 0.2 and 0.3, respectively, while, after the steep and uncertain $\Gamma = -2.1$ result for Cep OB5, there are three associations with $\Gamma = (-1.7, -1.7, -1.6)$ with errors of (0.3, 0.4, 0.3); for comparison, NGC 6913 has $\Gamma = -1.1 \pm 0.6$, i.e. near the mean yet very uncertain. Thus it is difficult to claim a trend of increasing error with increasing deviation from the mean. However Massey et al. point out that all the $\Gamma$ values are within $2\sigma$ of the mean value, suggesting to them that all the IMFs are the same within the statistical uncertainties. But given the large uncertainties in the $\Gamma$s, it seems just as likely that there is a real and significant spread in $\Gamma$, by at least 0.6. Such a difference, if real, would argue strongly against universality.

The situation is made more complicated by the fact that the lowest mass bins at 5–7 $M_\odot$ were omitted from all the fits due to probable incompleteness and the presence of pre-main sequence stars among these objects. But in 11 of 13 cases this omitted point lies *above* the fitted power law, so that if incompleteness was the dominant effect one would have to conclude that the IMF steepens in most cases below 10 $M_\odot$.

Massey et al. (1995) also summarize their previous results for massive stars in four LMC and one SMC clusters. The resulting values of $\Gamma$ range from $-1.1 \pm 0.1$ to $-1.6 \pm 0.2$. Again, whether or not this should be interpreted as evidence for universality or for a spread in $\Gamma$ is subjective, although it is certainly important that the range and mean are in agreement with the Milky Way results, suggesting that there is no strong dependence of the high-mass IMF slope on metallicity.

Oey (1996) has carried out a similar analysis for massive stars associated with six LMC superbubble HII regions and one LMC classical HII region. The derived values of $\Gamma$ for $m > 10$ range from $-1.0 \pm 0.1$ to $-1.8 \pm 0.4$. The steeper IMFs found by including stars outside superbubbles are attributed to contamination by field stars. The steepest value, $\Gamma=-1.8$, for D25, only extends out to about 20 $M_\odot$, and has a large uncertainty ($\pm 0.4$), but it is still significant that 4 of the 7 regions studied had $\Gamma=-1.6$ to $-1.8$. Once again, the spread is within $\sim 2\sigma$ of the errors, but an interpretation in terms of universality is questionable. Likewise, the (unweighted) mean $\Gamma$ is $-1.5$ for this sample, compared to the (weighted) mean of $-1.1$ for Milky Way groups, and a decisive statistical statement about the difference between the LMC and Milky Way cannot be made.

Deciding between IMF universality within the uncertainties and true IMF variations becomes even more problematic when one considers independent studies of the same region. For example, Oey & Massey (1995) derived $\Gamma = -1.3 \pm 0.2$ for the massive stars in the LMC superbubble LH47. A recent independent study of this region by Will et al. (1997) quotes a result $\Gamma = -1.2 \pm 0.1$ using the same evolutionary tracks as Oey & Massey, in apparent agreement. However, the Will et al. value was fit only between 2.5 and 30 $M_\odot$. They clearly state (p. 469) that the value of $\Gamma$ for masses between 12 and 85 $M_\odot$, roughly the same range as in Oey & Massey, is $-2.1$, much steeper than the Oey & Massey result.

Massey & Hunter (1997) recently estimated the IMF of the very dense cluster R136 in the 30 Doradus region of the LMC, using HST spectra for the most luminous stars. They derive a value of $\Gamma$ of $-1.3$ to $-1.4$ between 20 and 120 $M_\odot$, although it should be noted that spectroscopy was only possible for $m > 50$. Hunter, O'Neill et al. (1996) had found $\Gamma \approx -1$ for the intermediate-mass stars in R136, with no evidence for mass segregation. An earlier study of R136 by Brandle et al. (1996, not referenced in Massey & Hunter) combined adaptive optics imaging with HST data, using similar evolutionary tracks as in Massey & Hunter, but the IMF results were significantly different. For the entire field of view Brandle et al. find $\Gamma = -1.6 \pm 0.1$ for the mass range 6–100 $M_\odot$. For the central 0.4 pc they obtain $\Gamma = -1.3$, for 0.4 pc $<$ r $<$ 0.8 pc the result is $-1.6$, while for r $> 0.8$ pc they find a best fit with two power laws: $\Gamma = -1.5$ for 4–25 $M_\odot$, and $\Gamma = -2.2$ for 12 to about 50 $M_\odot$. Since the Massey & Hunter study apparently included stars out to large projected radii, the Brandle et al. IMF is considerably steeper, and the finding of strong mass segregation disagrees with Hunter, O'Neill et al. (1996).

It is admittedly difficult to know what to make of these differences between studies, since various details of the procedures (e.g. effective temperature scale, treatment of reddening) differ. The differences do suggest that systematic uncertainties are larger than the fitting uncertainties quoted in inidividual studies.

Despite the uncertainty concerning the question of universality, some of the conclusions of Massey et al. appear valid. In particular, *within the observational uncertainties* there is little evidence for a systematic trend of IMF slope with metallicity, galactocentric radius in the Milky Way, or with stellar density over a factor of about 100. Hunter (1995) shows $\Gamma$ for the Milky Way groups as a function of galactocentric distance and stellar density. A similar plot of $\Gamma$ versus galactocentric radius for intermediate-mass stars in the open clusters studied by Tarrab (1982) is given in Fig. 70 of S86. Trends can be seen in these plots, but the uncertainties are large enough that nothing definite can be said. This does *not* mean that the IMF is necessarily universal, only that the IMF probably does not vary with these properties, within the uncertainties. The question of the magnitude of the uncertainties remains, however, the fundamental one, and much more spectroscopy to deeper magnitudes and comparison with different sets of evolutionary tracks are required before an answer can be given. My impression of the situation is that either the systematic uncertainties are so large that the derived values of $\Gamma$ are relatively meaningless or else there are probably significant IMF variations for massive stars in different clusters and associations.

One particularly interesting result for massive stars is the finding that massive *field* stars in the LMC, SMC, and Milky Way display a very steep IMF, with $\Gamma$ between –3 and –4 (see Massey 1997, sec. 4.4). The observational issue remains somewhat confusing because such a radically steeper high-mass IMF should lead to a significant steepening of the LF, yet Bresolin et al. (1995) find no difference in LF between (field + associaton) and field stars alone in M101.

## 4. Intermediate-Mass Stars in Clusters

The "cleanest" place for estimating the IMF is the 1–15 $M_\odot$ range in star clusters, for a number of reasons. 1. If the cluster is older than about ten (or so) million years, the main sequence is fairly well defined, and problems due to the presence of pre- and post-main sequence stars are minimized. The IMF can be estimated from the LF and the m-L relation, without the uncertainties associated with counting between evolutionary tracks. Photometry alone is sufficient, without the need for spectroscopy that occurs both for more massive stars and for PMS stars in which circumstellar material can affect the colors. This conclusion is uncertain, however, since Belikov & Piskunov (1997) have concluded that PMS stars can introduce features in the LF for clusters somewhat older than $10^7$ yr. 2. The m-L relation for main sequence stars is believed to be reasonably well-understood in this mass range, without the uncertainties due to inflections that occur at smaller masses. For a recent comparison of empirical and theoretical m-L relations see Fig. 9 in Meusinger et al. (1996). A possible exception is the lingering uncertainty concerning the treatment of core convective overshoot. 3. The effects of unresolved binaries are probably not severe in this mass range compared to smaller masses (see below, however). 4. For clusters, the huge uncertainty due to the history of the galactic SFR that plagues the field star IMF around 1–1.5 $M_\odot$ (and larger masses if the SFR history has non-monotonic features) is absent. 5. The corrections due to the dependence of scale height on mass, which plays a strong role in the determination of the field star IMF, is also absent. 6. Cluster stars were all formed in the same relatively small volume of space, and so, unlike the field stars, give an estimate of a specific realization of the IMF, rather than some ill-defined ensemble average. For this same reason, clusters are not subject to the uncertainties in distances that can be a significant source of systematic error in estimating the field star IMF (Kroupa et al. 1993).

There are cluster-specific uncertainties which partly compensate for these advantages. These are summarized in S86 and Phelps & Janes (1993). The most important of these are the problems of contamination by field stars, completeness at small masses, the likelihood of radial mass segregation, perhaps even for the youngest clusters, and the small sample size. The first three of these problems are related, since in order to guard against radial segregation a large area must be surveyed, but the contamination by field stars becomes greater at larger distances from the cluster center. Since nearly all findings of radial mass segregation find higher-mass stars more concentrated than lower-mass stars, this effect means that the true IMF may be steeper than the derived IMF; i.e. derived values of $\Gamma$ must be considered as upper limits.

There have been a large number of studies of cluster IMFs for open clusters in the 1–15 $M_\odot$ mass range. Most work earlier than 1985 was reviewed in S86, where the LFs for 19 young and intermediate age clusters and associations taken from the literature were converted to IMFs using the same m-L relation. Several other estimates of open cluster IMFs were discussed and displayed. A particularly interesting paper is the study of 75 fairly young clusters by Tarrab (1982), whose results are displayed in various ways in Figs. 34, 35, and 70 of S86.

There have been major advances since that time. One of the most careful and important studies was by Phelps & Janes (1993; PJ) who estimated the IMF for eight clusters with ages 10–70 $\times 10^6$ yr. The IMF could be estimated in the range (1.2–1.9) to (7.9–18) $M_\odot$ for the various clusters. Following PJ, the very flat IMF ($\Gamma = -0.4$) found for NGC 436 should probably be excluded from discussion because of the small area surveyed and the likelihood that it is old enough that mass segregation has occurred. The remaining cluster IMFs have derived values of $\Gamma$ between –1.1 (NGC 663 and NGC 659) and –1.8 (NGC 581). In three cases (NGC 457, 581, and 663) extremely good power law fits were obtained. Two of these, NGC 663 ($\Gamma$ = –1.06) and NGC 581 ($\Gamma$ = –1.78) are shown in Fig. 1. Besides these power law fits, PJ detected gaps in the IMFs of at least three clusters that have low probability for occurring at random, using the criterion given in S86. One of these (NGC 103) is also shown in Fig. 1. This possibility of real structure in individual IMFs is extrmely important, and has been neglected in most other cluster studies, where dips and peaks in localized mass ranges are implicitly assumed to be due to counting errors.

A problem occurs when estimating the average value of $\Gamma$ for all the clusters, because of mass segregation. PJ find that, for all clusters (except NGC 436) combined, $\Gamma = -1.40 \pm 0.13$ in the mass range 1.4–7.9 $M_\odot$. However this weighted mean includes three clusters with relatively flat slopes which were not sampled to large areas. One of these clusters (NGC 663) contains a large number of stars and so contributes significantly to the mean. The cluster with the largest number of stars (NGC 457, 600 stars) yields $\Gamma = -1.37$, but in this case the photometry does not cover the entire area of the cluster, and for the two remaining clusters for which PJ state that the photometry probably does not cover a large enough area to detect low-mass stars (NGC 659 and Tr 1), the values of $\Gamma$ are –1.14 and –1.41. The two clusters surveyed with the large format CCD chip (NGC 103 and 7790) give $\Gamma = -1.47$ and $\Gamma = -1.67$. It is unclear whether the cluster with the steepest IMF (NGC 581, $\Gamma = -1.78$) was surveyed to large area, but if it wasn't the IMF should be even steeper.

Figure 1. Intermediate-mass IMFs for three well-populated (∼few hundred members) open clusters with similar ages from Phelps and Janes (1993). Best fit values of $\Gamma$ are indicated. NGC 581 and NGC 663 exhibit good power-law fits, but with very different indices. NGC 663 was not observed to large radii, and so the IMF may be steeper due to radial mass segregation. NGC 103 is not well-fit by a power law but instead shows plateaus and dips.

I conclude that, if mass segregation is important, then the best estimate of a representative $\Gamma$ is between –1.5 and –1.8 in the 1.2–10 $M_\odot$ mass range, not significantly flatter than the estimates for clusters given in S86. If mass segregation is insignificant for the clusters not surveyed at large areas, then the derived spread in $\Gamma$ (–1.1 to –1.8) is much larger than the uncertainty estimates for $\Gamma$ (0.05 to 0.24), estimates which include uncertainties due to field star fluctuations, which are larger than Poisson uncertainties. In this interpretation, the PJ results should be considered as evidence for a non-universal IMF, an interpretation strengthened by their finding of statistically unlikely gaps in several of the cluster IMFs.

A number of papers could be cited to support either view. For example, the Vazquez et al. (1997) UBVRI survey of the open cluster Cr 272 (age $\sim$ 13 Myr) gives $\Gamma = -1.8$ for 1.3–10 $M_\odot$. With the exception of one (low) point at 4.5 $M_\odot$, a power-law fit appears good. (However two of the points in their Fig. 7b do not match the values given in their Table 5.) Brown et al.'s (1994) photometric study of the Orion OB1 association gives, excluding large masses for which spectroscopy is needed, $\Gamma = -1.8$ for the mass range 4–15 $M_\odot$ for subgroup a (their Fig. 15a); similar values were found for the other two subgroups studied. Subgroup a is probably old enough ($\sim$11 Myr) that it is unlikely that contributions from PMS stars are significant at masses as large as 4 $M_\odot$, but this is a problem for the other (younger) subgroups. A notable feature of the Brown et al. study (besides the interesting discussion of energy output and ISM interaction) is their use of a Kolmogorov–Smirnov test to estimate the probability of a given $\Gamma$, making it clear how uncertain the results are. For example, demanding a confidence level of 70%, power laws with $\Gamma$ between –1.3 and –2.1 are consistent with the observations for Ori OB1a.

However a number of recent cluster studies can be cited as evidence for strong variations in the IMF at intermediate masses. Meusinger et al.'s (1996) proper motion study of the Pleiades cluster covering a large area (shown in Fig. 2) gives a flat ($\Gamma \approx 0$) IMF below 1.1 $M_\odot$, but between 1.1 and 3.5 $M_\odot$ the IMF is well-fit by steep power law with $\Gamma = -2.3$. Note that this is one of the most thorough studies of any open cluster yet published. At the flat end of the spectrum of IMF results for this mass range, the IMF of the $\rho$ Oph cluster (Williams et al. 1995, see below), while more problematic because of the deeply embedded nature of the sources, gives $\Gamma \approx 0$ between 0.1 and 4 $M_\odot$! A study of the intermediate-mass IMF in the central regions of the young cluster Berkeley 86 by Deeg and Ninkov (1996) gives $\Gamma = -1.3 \pm 0.3$ over 1.1 to 20 $M_\odot$, but the IMF presented (their Fig. 9) is much better fit by two power laws, with $\Gamma \approx -0.6$ for 3–16 $M_\odot$ and a very steep $\Gamma \approx -2.7$ between 1.2 and 3.5 $M_\odot$, one of the most distinctively "non-universal" IMFs yet found. It should be noted, however, that since only the core of the cluster was studied, the effects of mass segregation are unknown, and that there were only 19 objects in the mass bins above 3 $M_\odot$.

Space requirements preclude a discussion of several other recent open cluster studies (e.g. Ninkov et al. 1995, Aparicio et al. 1993, Patat & Carraro 1995), except to say that each of them have their own idiosyncrasies which demand that the reader actually study the papers before accepting the conclusions given in the abstracts.

Figure 2. IMF of the Pleiades cluster, from Meusinger et al. (1996). The points at low masses represent various combinations of evolutionary models and conversions to the color-magnitude diagram. Solid curve is weighted average IMF, dashed curve is result for "best" model (see Meusinger et al. for details). A correction for unresolved binaries with a binary fraction of 0.4 was made using a Monte-Carlo simulation. The IMF shows a sudden transition at $m \approx 1$ from $\Gamma \approx 0$ to a steep power law with $\Gamma \approx -2.4$ at larger masses.

A large amount of recent work has been devoted to the IMF of 1–15 $M_\odot$ stars within moderately young clusters and associations in the LMC. Much of this work has been aimed at resolving the large discrepancy between earlier studies by Mateo (1988), who found very steep IMFs ($\Gamma \sim -2$) and Elson et al. (1989), who found very shallow IMFs ($\Gamma \sim 0$). A major study of this problem was given by Sagar & Richtler (1991; SR), who estimated intermediate-mass IMFs for five LMC clusters from color-magnitude diagrams. The data contain a sufficiently large number of cluster members (230 to 600) that a reasonably large number of mass bins could be included without large Poisson uncertainties. Besides treating different methods for completeness corrections, SR also discuss the effects of unresolved binaries (see below), different sets of evolutionary tracks, a search for radial segregation, and they present Monte Carlo simulations to estimate the errors due to various effects. The results for their preferred completeness corrections (their $x_2$ in Table 8), vary from $\Gamma = -1.0\pm0.1$ to $-1.3\pm0.3$ for four clusters, for their inner "Ring 1." The mass ranges covered are from (2.1–2.6) $M_\odot$ to (5.7–12) $M_\odot$. The results for their outer "Ring 2," for five clusters, vary from –0.8±0.3 to –2.2±0.3. Although SR dismiss the differences as being within the statistical uncertainties, it seems highly significant that all the Ring 2 IMFs are steeper, by 0.2 to 1.1 in $\Gamma$. The Ring 2 IMFs are noisier (see their Fig. 4) and may be more affected by uncertainties in completeness corrections but still this difference suggests that the average $\Gamma = -1.1$ they prefer must be considered an upper limit. Another important result is that the derived IMFs are steeper for other choices of evolutionary tracks, as steep as –1.9 for NGC 1711. The average values of $\Gamma$ for "ring 1" for all three sets of tracks range from –1.05 to –1.5, with a mean of –1.27. Considering the lack of basis for choosing between evolutionary models, this value of $\Gamma$ seems like a more reasonable representation of their results for "Ring 1."

SR present a very useful investigation of the effects of unresolved binaries, using simulations in which stars are paired at random for an IMF with a given $\Gamma$. The results indicate that the true IMF will be considerably steeper if the binary fraction is significant. Their Table 10 shows that their preferred $\Gamma \approx -1.1$ implies a "true" $\Gamma$ of about –1.5 if the binary fraction is greater than about 0.4. The result is a little steeper if the $\Gamma \approx -1.3$ result (averaging over sets of models) is used. SR also show that the magnitude of the effect depends sensitively on the index of the true IMF, being larger for flatter IMF, as expected. The binary fractions in these clusters is of course unknown, but it seems unlikely that dynamical evolution can significantly reduce the binary fraction for such young clusters (see the N-body simulations by Kroupa 1995c and de la Fuente Marcos 1997).

It therefore seems that, on the basis of both the radial segregation and unresolved binary results of SR, as well as the dependence on evolutionary models, the average $\Gamma$ for these clusters is probably significantly steeper than –1.1, by perhaps 0.3 to 0.9. This conclusion should be borne in mind in the following discussion of later work on LMC clusters.

Most recent results have confirmed the conclusions of Sagar & Richtler (1991), that the derived IMF is extremely sensitive to different types of data and analysis methods, particularly the treatment of completeness corrections, and that the "correct" value of $\Gamma$ (if indeed the IMF is a power law) lies *somewhere* between these two extremes found by Mateo and Elson et al. I discuss a few of the relevant papers here, excluding studies that concentrated on relatively massive stars because of the need for spectroscopy at large masses, as discussed above.

Subramaniam & Sagar (1995) estimated $\Gamma$ for four young LMC clusters using three different sets of evolutionary tracks, and including the effect of unresolved binaries. For a binary fraction of 30%, they find a range of $\Gamma$ between –1.35 and –1.90 for their "Model 1," and –1.1 to –1.35 for their "Model 3." This suggests that the evolutionary model uncertainties may be fairly large, up to 0.55 in $\Gamma$ (for NGC 1711) although the authors show that it may be possible to choose preferred evolutionary calculations by matching the morphology of the color-magnitude diagram.

Vallenari et al. (1993) estimated the IMF for stars in NGC 1948 (in the super-shell LMC 4) for masses $m = 3.5 - 12$. Their results show how the derived $\Gamma$ depends on the adopted (and uncertain) age, whether or not evolutionary models with significant convective overshoot are employed in the m-L relation, and whether or not a correction for unresolved binaries is included. Different combinations yield $\Gamma$ between –0.8±0.3 to –1.3±0.3. Will, Vázquez et al. (1995) found $\Gamma = -1.2 \pm 0.1$ for 2.5–20 $M_\odot$, or $\Gamma = -1.1 \pm 0.1$ for 1.6–20 $M_\odot$ for the assocation NGC 1962-65-66-70. Bencivenni et al. (1991) found $\Gamma = -1.4$ for 2–20 $M_\odot$ in the NGC 2004 cluster, although because they only compare the observed and predicted cumulative LFs, this $\Gamma$ only applies at the lower end of the mass range, where crowding and completeness corrections are important. Subramaniam & Sagar (1995) find $\Gamma = -1.25$ to –1.65 for this cluster, depending on the adopted evolutionary models, for a binary fraction of 30%, in good agreement.

Banks et al. (1995) studied the LMC cluster NGC 2214. An attempt to fit a single power law over the mass range 1.2–6.3 $M_\odot$ gave $\Gamma \approx -1.1$. However their IMF rather clearly shows a change of slope at around 3 $M_\odot$ (see their Fig. 14), and they find $\Gamma \approx -0.8$ for $m < 3$ and $\Gamma \approx -2$ for $m > 3$, after attempting to correct for an artifact due to a saturation effect. This illustrates the danger of assuming a single power law even over such a narrow mass interval, and that the IMF may be steeper at $m \gtrsim 3$ than given in previous papers. Banks et al. note that the same sort of steepening can be seen in some of the results of Sagar & Richtler (1991). Subramaniam & Sagar (1995) also studied NGC 2214 and give estimates for $\Gamma$ between –1.10 and –1.35, depending on the adopted evolutionary models, for a binary star fraction of 30%, in good agreement with the single power-law fit of Banks et al. However their Fig. 6d indicates that a steeper IMF is required for $m \gtrsim 2$, supporting the above conclusion.

An interesting case study is NGC 1818, which has been recently studied independently by two groups. Will, Bomans, & deBoer (1995) found a good power-law fit with $\Gamma = -1.1$ to –1.3 (with fitting uncertainties of ±0.1) for the mass range about 1.5–8 $M_\odot$, depending on which of four evolutionary tracks were used. The lack of sensitivity to the adopted models is encouraging. Hunter at al. (1997) used HST data for this cluster, resulting in less severe crowding effects. They were able to estimate the IMF in the 1–8 $M_\odot$ range (they quote a mass range based on the lower and upper limits of the lowest and highest mass bins) and find a very good power law fit with $\Gamma = -1.2 \pm 0.1$. One slight caveat is that Hunter et al. included their lowest mass bin result, which appears to suffer from incompleteness, in their fit; the appropriate $\Gamma$ may be somewhat steeper if this point was omitted, although apparently not by much. The IMF will be slightly flatter ($\Gamma = -1.0 \pm 0.2$) if older isochrones had been used. Hunter at al. find no evidence for mass segregation. The agreement with Will et al. (1995) is very good, especially considering the differing observational data and techniques for correction for crowding, completeness, and field star contamination. This suggests that the IMF for NGC 1818 is relatively well-determined, except for the possible effects of radial segregation and unresolved binaries, as described above.

Figure 3. Combined intermediate-mass IMF for the associations in the Magellanic Cloud bridge studied by Grondin et al. (1992). Solid line is a least squares fit with $\Gamma = -1.3$. However the IMF shows rather clear evidence for two or more power-law regimes.

Grondin et al. (1992) derived the IMF in the 1.5–12 $M_\odot$ mass range from the combined LF of five small associations in the bridge of the Magellanic Clouds. The total number of stars in the combined sample was 214, after subtraction of the Galactic contamination correction and omission of the two most luminous bins (too few stars) and the faintest bin (incomplete). With this many stars it was possible to sample the 1.5–12 $M_\odot$ mass range with a fairly large number of mass bins. The result is shown in Fig. 3. The solid line is a straight least-square fit, giving $\Gamma = -1.3 \pm 0.1$. However Fig. 3 suggests that, while this is a good fit to the four highest-mass bins, there is strong evidence for a well-defined flatter region with $\Gamma \approx -0.8$ between 2.3 and 6.4 $M_\odot$ and a very steep portion at low masses, with $\Gamma \approx -3$ for 1.4 to 2.5 $M_\odot$. As in other cases where possible multiple or non-power law structure appears, the assumption of a single power law IMF may tell us nothing about the IMF, in effect smoothing over the most important features. An important procedure should be the use of a statistical test, like the Kolmogorov-Smirnov test, to provide an estimate of the significance of a single power-law fit.

The demonstration of Massey et al. that colors alone cannot be used to construct a theoretical H-R diagram and estimate IMFs of stars more massive than about 15–20 $M_\odot$ would invalidate the IMFs derived for 14 LMC and SMC OB associations by Hill et al. (1994). Hill et al. attempt to argue that the problem is not significant, partly on the basis of similarity of their values of $\Gamma$ with those obtained using spectroscopy. However it is clear that the photometrically derived mean $\Gamma(\approx -2)$ for $m > 10$ is about 0.7 steeper than the (mostly) spectroscopically-derived mean $\Gamma$, just as was claimed should occur in one SMC association if the problem is severe by Massey et al. (1989). Unfortunately the problem apparently cannot be corrected for by simply adding some constant amount to each photometrically derived $\Gamma$. Still, the fairly small dispersion of LF slopes found in a number of galaxies (see Bresolin et al. 1996, Malumuth et al. 1996 for references) probably rules out severe IMF variations for high-mass stars, although "severe" remains to be quantified.

If the mass bins above 20 $M_\odot$ are omitted, there are still bins that cover lower masses down to about 2.7 $M_\odot$ in the study of Hill et al., so one might hope to learn something about the intermediate-mass IMF from this work. Hill et al. only fit power laws above 10 $M_\odot$, partly because there is an apparent change of slope at that mass (flatter at lower masses) and partly because incompleteness and field star contamination become severe. They state that these uncertainties should not have a large effect on the two bins at 6 and 8 $M_\odot$, although a basis for this statement is not given. If it were true, then there would still be four mass bins between 6 and 13.5 $M_\odot$, or five if the 20 $M_\odot$ bin is included. I refrain from giving values of $\Gamma$ in these ranges because incompleteness can be seen to set in at 10 $M_\odot$ and 8 $M_\odot$ in two of the clusters, and so it seems likely that incompleteness corrections will dominate the counts for these mass bins. Similar remarks apply to the study of the giant H II region NGC 595 in M 33 by Malumuth et al. (1996).

Hill et al. (1995) have derived IMFs for the LMC associations LH 52 and LH 53 in the mass range 4–20 $M_\odot$. For LH 52 they find $\Gamma = -1.1 \pm 0.1$, but for LH 53 they find $\Gamma \approx -2$, with the precise value depending on the assumed age. With seven mass bins used in each fit, there is certainly an apparent difference in slope between the two regions, although it has to be remembered, as in all these LMC and SMC studies, that the derived IMFs below 8–10 $M_\odot$ are dominated by the large and uncertain completeness/contamination corrections.

Combining these results with the study of LMC clusters by Sagar & Richtler (1991), who found an average $\Gamma \approx -1.3$ (averaging over different models and ignoring the steeper IMFs found in their larger ring), and ignoring the discrepant results discussed above, suggests that the IMF in the 1–10 $M_\odot$ is relatively shallow and constant (although the Banks et al. result for $m > 3$, the Grondin et al. 1992 result for 1.4–2.5 $M_\odot$, and the Hill et al. 1995 result for LH 53, should be noted). The result $\Gamma \approx -1.3$ is significantly flatter than was inferred for the typical values in Galactic open clusters discussed above, where I suggested that $\Gamma \approx -1.5$ to –1.8 might be representative, although with notable exceptions.

A major problem with this conclusion concerns the effect of unresolved binaries. Hunter et al. (1997) estimated the effect for NGC 1818 by assuming that half of the stars in each mass bin was of equal luminosity and mass, and found that this steepened the IMF slope by roughly 0.15–0.2. However, the effect of binaries depends on the assumed distribution of mass ratios, and also requires that the contributions from the *entire* IMF be included, not just the mass range accessible, since systems with luminosities above the studied range will contribute secondaries to this range if mass ratios different from unity are allowed, and the effect becomes more important for shallower IMFs.

Even without considering masses outside the studied range, the simulations of SR discussed above, which assumed random pairing of masses, suggest that these values of $\Gamma$ should be steepened by about 0.3 if the binary fraction exceeds about 0.4. Combined with the possibility of radial mass segregation, this suggests that, if the LMC clusters can be fit by a single power law, it may have an index at least as steep as $\Gamma = -1.5$. Of course the Galactic open cluster IMFs should also be steepened due to unresolved binaries, although the effect will not be as great because the apparent IMFs are steeper. Thus, just as in the case of massive stars, it not possible to say at present whether or not there is a significant difference between the IMFs of Galactic and LMC clusters, although the difference, if it exists, is likely at the level $\Delta\Gamma \sim 0.3$.

## 5. Pre-Main Sequence Stars

The development of near-infrared array cameras in the past decade has allowed a census of stars within very young stellar groups that are still embedded in their parent interstellar gas and dust. An extremely useful review of this area is given by Zinnecker et al. (1993), who discuss a large number of earlier studies back to the pioneering paper of Wilking & Lada (1983) on the K-band LF of the $\rho$ Oph cluster. Unfortunately there are severe problems involved with estimating IMFs from such data, especially if one attempts to constrain the IMF from a near-IR LF (the K-band LF is almost always used for this purpose). A fundamental problem is that pre-main sequence stars are generally not evolving horizontally in the H-R diagram, so a cut at constant K (say) magnitude samples stars with a variety of masses: there is no one-to-one relation between luminosity and mass. Instead, the number of stars in a given magnitude interval is an integral over the IMF (and the SFR history), and the form of the resulting LF is less sensitive to the form of the IMF. This insensitivity can be seen clearly in the explicit mathematical formulation presented by Fletcher & Stahler (1994a,b); in particular see Fig. 11 of Fletcher & Stahler (1994b). The problem is similar to that encountered in trying to use the visual LF to constrain the IMF of massive main sequence and post-main sequence stars. Attempts to circumvent this problem by using an effective mass-luminosity relation derived from evolutionary tracks (Zinnecker et al. 1993) is probably inadequate for constraining the form of the IMF, especially since it effectively requires that the stellar group be characterized by a single isochrone. The K-band LF *can* be used to constrain another integral relation. The dependence of the K-band LF on the IMF illustrated in Giovannetti et al. (1997, Figs. 10 and 11) is largely of this nature. A popular integral constraint is the ratio of the number of stars above and below a particular mass (say 1 $M_\odot$), and this is of course an important constraint, if it can indeed be estimated. Reasonably modest attempts to use the near IR LF of embedded

PMS to constrain the high/low ratio are the studies of Megeath et al. (1996), Meyer (1996), and Carpenter et al. (1997), where references to earlier work may be found. Unfortunately there is a tendency for KLF papers to compare this ratio with the field star ratio, which is very poorly determined. (Even worse, they usually compare with the MS 79 IMF, which is entirely superseded.) Meyer (1996) recognized the need to compare results for several clusters rather than test each against the field star IMF, and gives a useful summary (his Table 6.1 and Fig. 6.1) for six groups. However, in order to derive the *form* of the IMF for PMS stars, it is necessary to count stars between evolutionary tracks for models of different masses.

Another problem with the use of the LF is that it may be strongly affected by the SFR history of the region under consideration (again, see the formulation in Fletcher & Stahler 1995a,b). We know essentially nothing about this SFR history; it might be roughly constant, or a monotonically increasing or decreasing function, or series of bursts, and it might be different for different masses. This is the same problem that precludes the derivation of the field star IMF for masses corresponding to main sequence lifetimes near the galactic main sequence turn-off mass. Counting stars in the H-R diagram avoids this difficulty, because comparison with evolutionary tracks can in principle give the distribution of ages and masses separately.

The problem is made even worse because the rate of evolution along PMS tracks is affected by various physical effects (e.g. deuterium burning), leading to sizeable features in the LF that have nothing to do with the IMF (see Zinnecker et al. 1993, Piskunov & Belikov 1996, Belikov & Piskunov 1997). Extinction uncertainties and clumpiness will also mask the form of the IMF derived from the K-band LF, unless the LF is exactly a power law (Megeath 1995; Comeron, Torra, & Rieke 1996). This appears to be a serious effect for the deeply embedded clusters to which the KLF approach has been applied, and will even alter integral constraints such as the high/low ratio.

The determination of IMFs from counting stars in the H-R diagram still has its own problems. The relation between effective temperature and color is apparently not degenerate, as was the case of massive stars, but the effective temperature and bolometric correction scales still depend on uncertainties associated with molecular opacities and convection in model atmospheres and evolutionary tracks (see Luhman & Rieke 1997, Allard et al. 1997). Ideally, optical and near-IR spectroscopy are needed to place the PMS stars in the H-R diagram, especially to circumvent effects associated with veiling and infrared excesses (e.g. Strom & Strom 1994, Hillenbrand 1997, Luhman & Rieke 1997, Carpenter et al. 1997, Alcala et al. 1997), but then the sample size is reduced compared to photometric samples. The effect of accretion on the positions of objects in the H-R diagram is an additional source of uncertainty, since accretion is believed to dominate the luminosity of low-mass PMS stars (e.g. Fletcher & Stahler 1994a,b).

Establishing membership is a problem unless proper motion and radial velocities are available, and radial mass segregation is probably present at birth (e.g. Hillenbrand 1997; see, however, Carpenter et al. 1997), requiring surveys over a relatively large area in which contamination by field stars becomes increasingly severe. Finally, most of the available studies concentrate on the low-mass IMF, where the effects of unresolved binaries will be most severe. It should be emphasized that the magnitude of the effect of unresolved binaries cannot be estimated simply by assuming some fraction of binaries and mass ratio distribution for the observed sample; instead it is necessary to simulate the effect by including an entire IMF of various assumed forms, since, unless the IMF is relatively steep, a lot of low mass stars can be missed because they are paired with more massive objects. (Kroupa et al. 1991). Clearly, assuming a mass ratio of unity could dangerously underestimate this effect, and would simply adjust the mass assigned to a given object.

Despite all these problems, there are several interesting and suggestive recent results. For the $\rho$ Oph cluster, Williams, Comeron et al. (1995) showed that two independent previously-used methods for estimating masses from near-IR colors (Comeron et al. 1993, Strom et al. 1995) agree surprisingly well, and showed how spectra of six of the objects can be used to derive upper mass limits for the brown dwarf candidates and to validate some of the assumptions of the photometric method. Combining the two studies in order to reduce the noise, Williams et al. present an IMF that is flat ($\Gamma \approx -0.1$) well into the substellar range. They argue that further accretion of mass is unlikely for the types of sources in the sample, so this IMF should be representative of a "finished" cluster. This flat (in log m) very low-mass IMF agrees well with several other studies, and is not inconsistent with what Kroupa et al. (1993) and Reid & Gizis (1997) found for field stars, although the Mera et al. (1996) low mass field star IMF is steeper at very low masses, as discussed above. Unresolved binaries might steepen the low-mass end, as might radial segregation, but the magnitude of the effects are extremely uncertain. Kroupa (1995c,d) has discussed in detail the effect of unresolved binaries on the luminosity function of N-body cluster simulations. Unfortunately it is not clear how to compare his results (or the earlier results of Kroupa et al. 1991 or the analytic results of Piskunov & Malkov 1991) with the IMFs derived from counting pre-main sequence stars between evolutionary tracks, although certainly the direction of the effect will be the same, i.e. to make the derived IMF flatter than it really is. Furthermore, the degree of the correction depends sensitively on the assumed form of the distribution of mass ratios.

Perhaps the most interesting result of Williams et al.'s study, not pointed out by the authors, is the fact that the IMF per unit log mass is basically flat *all the way out to 4 $M_\odot$*! There are about 10–15 stars in each of the 3 highest mass bins (at 0.7, 2, and 4 $M_\odot$), so a Poisson fluke is unlikely. Either unresolved binaries and/or mass segregation, or some other systematic effect, are making the results meaningless (and it is unlikely that unresolved binaries have much effect above 1 $M_\odot$), or we have here outstanding evidence for a non-universal IMF. There are apparently no other studied clusters with IMFs this flat at such (relatively) large masses. The only example that comes close is NGC 436 in the Phelps & Janes (1993) cluster study, which had an apparent $\Gamma \approx -0.4$ in the 1.5–5 $M_\odot$ mass range; however Phelps & Janes excluded this cluster because of the likelihood of mass segregation and the small area surveyed. Comparing this $\Gamma \approx 0$ result with the $\Gamma \approx -2.6$ result for the same mass range in the Pleiades cluster (Meusinger et al. 1996) illustrates the very real possibility of radical IMF variations.

An improvement in methodology for extracting an IMF from multiband near-IR photometry has been given in Comeron, Rieke, & Rieke's (1997) study of the low mass IMF in NGC 2024, an HII region located in the Orion B giant molecular cloud. This is a particularly interesting region because of the repeated suggestion that it is the product of star formation triggered by the neighboring Orion OB1b subgroup. Rather than assume coeval star formation and a single isochrone, as in most previous work, Comeron et al. use a probabilistic technique that samples a mass for a range of possible ages of each star. Repetitive sampling for each object then gives an estimate of the most probable mass, and, for the cluster, the most probable IMF. This approach allows for star formation over an extended time and yields error bars that include the age uncertainties. The resulting IMF does depend on the assumed maximum age, but the overall shape is not severely affected.

The resulting IMF per unit log mass is relatively flat between 0.06 and 1 $M_\odot$, with $\Gamma \approx -0.2$. A reconsideration of the $\rho$ Oph IMF using the same method gives $\Gamma \approx -0.4$ in the same mass range (the data for higher-mass objects were not included). Comeron et al. note that $\Gamma \sim 0$ for the low mass IMF has also been found in well-studied clusters such as the Hyades, Pleiades, and Praesepe (see Comeron et al. for references, and the discussion below for Pleiades and Praesepe), and agrees with most estimates for the field star IMF.

Luhman & Rieke (1997) have carried the study of embedded PMS populations further, by using low-resolution K-band spectroscopy to obtain spectral types for PMS stars in the L1495E cluster within the Taurus cloud complex. Luhman and Rieke give a thorough discussion of the uncertainties associated with the effective temperature scale, the determination of bolometric luminosities, and the choice of theoretical evolutionary tracks. The spectroscopic sample for L1495E is incomplete for faint sources, and JK imaging was used to estimate completeness corrections. The resulting IMF depends somewhat on the adopted models and incompleteness corrections, but the results suggest that the IMF per unit log mass is relatively flat or slightly increasing toward lower masses for m $\lesssim 0.4$, with no turnover, at least for their favored evolutionary tracks. From m $\approx 0.4$ to 1.5 the IMF agrees (probably fortuitously) with the S86 field star IMF.

Luhman & Rieke point out that the similarity in low-mass IMFs found for several embedded clusters, as well as for nearby unobscured but older clusters (Hyades, Pleiades, Praesepe, see below), suggests that the low-mass IMF is flat and does not vary much between regions differing by a factor of at least 100 in stellar density. Meyer (1996) found a tendency for the high (1–10 $M_\odot$)/low (0.1–1 $M_\odot$) mass ratio to increase with increasing stellar density for six young regions. If this is correct, and the low-mass IMF is relatively constant, then the intermediate-mass IMF must flatten with increasing stellar density. However the trend is not confirmed in the two higher-density clusters Mon R2 (Carpenter et al. 1997) and the Trapezium (Hillenbrand 1997). Still, the variation in the high/low ratio by a factor of five among these eight clusters suggests IMF variations above 1 $M_\odot$.

Almost none of the recent studies of any of these regions shows evidence for a low-mass turnover (a notable exception is Orion–see below). Studies of groups of PMS stars that do find evidence for a turnover at low masses usually suffer serious incompleteness at luminosities corresponding to the turnover mass

Figure 4. IMF for over 900 optically visible stars in the Orion Nebula Cluster from Hillenbrand (1997) for one choice of evolutionary tracks. Another choice of tracks shows the same form but is compressed toward somewhat larger masses (see Hillenbrand's Fig.15). The IMF above 0.3 $M_\odot$ appears as two steep segments separated by a plateau or dip, and shows a strong turnover below about 0.2 $M_\odot$.

(e.g. Alcala et al. 1997 for the Chamaeleon regions). However there are still large uncertainties associated with the results for all the embedded clusters. For example, it will be extremely difficult to disentangle the IMF from the effect of the large and highly irregular extinction (see Megeath 1995, Comeron, Torra, & Rieke 1996) unless the dependence on photometry is eliminated or extremely high-resolution submillimeter maps of column density become available for individual embedded clusters. One way around this and other problems is to concentrate on regions with large numbers of optically visible stars.

The most comprehensive study of a young star-forming region published at the time of this writing is Hillenbrand's (1997) study of the optically visible stars in the Orion Nebula Cluster (ONC, about 3 pc in radius centered on the Trapezium within Ori Ic). New V and I photometry and 5000–9000 Å spectroscopy were combined with existing data to construct an H-R diagram for nearly 1000 stars. This sample is about ten times larger than in the studies of embedded clusters described above. Foreground star contamination is treated using the results of several proper motion surveys (not possible in the deeply embedded case), and spectroscopic completeness is large and fairly constant with mass.

The IMF, reproduced in Fig. 4, was constructed by counting between masses for different isochrones in the H-R diagram, using two different sets of evolutionary tracks. Even though nearly 1000 stars were placed in the H-R diagram, there are only a small number of intermediate- and high-mass stars (about 6 stars with $m > 10$, about 14 stars with $m > 4$). For both sets of tracks, the IMF rises fairly continuously from large masses down to a few $M_\odot$, but then flattens down to about 1 $M_\odot$, then rises sharply again, finally turning over strongly at m $\sim$ 0.3. The exact value of the turnover depends on the evolutionary models, but the general form is the same for both sets of models. Adopting a cooler temperature scale would broaden the IMF below the peak, but leaves the general form unaffected. Comeron et al. (1997) claim the turnover in Hillenbrand's IMF is near the completeness limit, but the situation is far from clear; Hillenbrand presents arguments for the completeness of the sample well below the turnover. Deeper near-IR photometry will be needed to confirm this feature. In addition, a detailed study of the effects of unresolved binaries is required. If this turnover is real, it is in strong contrast to the flat IMFs found in several other clusters, and would mean that the low mass IMF is not universal (or there are large systematic uncertainties in the studies).

If a single power law is forced throug the IMF between 0.3 and 10 $M_\odot$, the resulting $\Gamma$ is around –1.4. However there appears to be a real deficit of stars in the mass range 0.5 to 2 $M_\odot$ within the ONC cluster compared to the field star IMF and several other well-studied clusters. Furthermore, independent of comparison with field stars, there is definitely a "ledge" between 0.7 and 2 $M_\odot$ that is unlikely to be a Poisson artifact. This is reminiscent of the flat IMF between 0.7 and 4 $M_\odot$ found in the smaller sample of Williams et al. for $\rho$ Oph, except that the derived $\rho$ Oph IMF stays flat to well below the hydrogen-burning limit, while the Orion IMF rises steeply just below 0.7 $M_\odot$, and then turns over at even lower masses.

Hillenbrand points out that there is strong evidence for mass segregation in the ONC, with the more massive stars near the center, reminding us that even young cluster IMF determinations must be suspected of finding too flat an IMF compared to the IMF that would be determined if a larger area could be sampled. Note however that Carpenter et al. (1997) do not find evidence for segregation at $m \lesssim 2$ in their study of the Mon R2 cluster, although higher-mass stars are clearly found to prefer the cluster center.

There have been several other studies, primarily photometric, of the *intermediate-mass* IMF for optically visible stars in very young clusters dominated by PMS stars. An example is the study by Ninkov et al. (1995) of the OB cluster in the HII region IC 1805 (within Cas OB6 within the molecular cloud W4). The result for the stars with either proper motions, spectra, or reddening consistent with membership is $\Gamma = -1.4 \pm 0.2$, between 3.5 and 21 $M_\odot$ (centers of log mass bins), but this is based on only 42 objects in four mass bins. This is still a significant result however, and in fact includes a larger number of stars than Hillenbrand's ONC IMF over a similar mass range. The situation is similar to that for studies of very massive stars in Milky Way and LMC clusters and associations, where the small sample sizes result in only a few usable mass bins.

It has been mentioned that most work on the *low-mass* IMF of PMS stars in embedded clusters yields a flat or slightly increasing IMF, in agreement with studies of nearby unobscured clusters like the Pleiades, Hyades, and Praesepe, and perhaps with the field star IMF, suggesting universality at small masses. Even the well-studied unobscured clusters, however, are still affected at very low masses by PMS stars and the same uncertainties concerning the effective tem-prature scale and evolutionary tracks apply to them as to the deeply embedded clusters. I do not attempt to review all the relatively recent work on the IMF of PMS stars in nearby unobscured clusters, although the thoroughness with which Meusinger et al. (1996) treat the problem for the Pleiades cluster (e.g. employing 30 combinations of evolutionary track and conversions between observational and theoretical color-magnitude diagrams), discussed in sec. 4, is worth noting as a model for future studies (see Fig. 2). The Meusinger et al. result $\Gamma \approx 0$ between 0.1 and 1 $M_\odot$ is consistent with both the $9\pm4$ detections in the 0.04 to 0.25 $M_\odot$ range by Williams et al. (1996) and the large-area brown dwarf search by Festin (1997), who found that $\Gamma$ is probably positive for $m < 0.05$, i.e. that the IMF F(logm) decreases with decreasing mass at extremely small masses.

Another well-studied cluster, particularly well-positioned at relatively high latitude, is Praesepe. The deep VIK photometry of Williams, Rieke et al. (1995), which should be complete down to a magnitude corresponding to $m \approx 0.08$, allowed the detection of six very faint sources. These sources were grouped into a single mass bin and the result combined with the LF from the detailed Jones & Stauffer (1991) Praesepe study at higher luminosities, using the $M_V - m$ relation of Henry & McCarthy (1993), to construct an IMF between 0.2 and 1.4 $M_\odot$. Although they quote $\Gamma = -1$, the index between 0.2 and 0.6 $M_\odot$ is about –0.3, while above 0.6 $M_\odot$ the IMF is much steeper with $\Gamma \approx -1.2$, as shown in their Fig. 7. The three bins above 1 $M_\odot$ are fit with $\Gamma \approx -1.7$. This relatively flat IMF at very low masses and rather sudden steepening above 0.6–1 $M_\odot$ is very similar to the results for L1495E and the Pleiades, discussed above, and to the field star IMF from either Kroupa et al. (1993) or Reid & Gizis (1997) for the low-mass stars, and S86 for the 1.2–4 $M_\odot$ range, although the steepening for the Pleiades ($\Gamma \approx -2.4$) and the field stars ($\Gamma \approx -2.7$) is more severe.

Thus the unobscured open clusters, embedded clusters, and field stars show rough agreement, $\Gamma \approx 0$ to –0.5, in the 0.1 to 1.0 $M_\odot$ mass range, although all the results are seriously dependent on uncertainties in the m-L relation, the models, radial segregation, unresolved binaries, etc.

The severity of the uncertainty at very low masses can be seen from studies of old open and globular clusters, where it can be assumed that even the lowest-mass stars are on the main sequence, so that the IMF can be derived directly from the m-L relation. A number of recent such studies (see von Hippel et al. 1996 and references therein) have found that even in these rather optimum cases it is extremely difficult to estimate the low-mass IMF, because the observed form of the LF is driven mainly by the m-L relation, not the IMF. Claims about the form of the IMF at small masses, whether for field stars or in clusters, therefore need to be regarded with caution.

An important point is that significant dynamical evolution is expected even in open clusters such as the Hyades and Praesepe. Such evolution can result in alteration of the present-day mass function as compared to its initial form, and detailed case-by-case N-body simulations would be needed (as in Kroupa 1995c,d). Much of the advantages of cluster IMFs over field stars is therefore lost for clusters older than about $5 \times 10^7$ to $10^8$ yr (depending on the mass and size of the cluster).

## 6. Galaxies Beyond the Magellanic Clouds

The IMF in galaxies more distant than the Magellanic Clouds remains effectively inaccessible to star counts except to say that the similarity in LFs for the brightest stars in several galaxies seems to rule out severe variations (e.g. Bresolin et al. 1996), where the meaning of "severe" is currently unknown. For high-mass stars it is not yet possible to obtain spectroscopy for a significant number of stars. For intermediate-mass stars, for which photometry could in principle be used to infer masses, the lowest masses that can be reached are so large that the accessible mass range is only a factor of two or three, and the uncertainties at these masses are very large. Recent attempts in this direction have been made by Hunter and coworkers using HST photometry. Hunter et al. (1996a) studied the M31 OB association NGC 206 and found $\Gamma = -1.4 \pm 0.5$ using four mass bins between 6.8 and 13.5 $M_\odot$ (center of mass bins). However the IMF presented does not resemble a power law, and so it is difficult to interpret the uncertain result. A similar study of NGC 604 in M33 (Hunter et al. 1996b) finds $\Gamma = -1.6 \pm 0.7$ for four mass bins between 7.5 and 16 $M_\odot$. The presented IMF could be interpreted as evidence for a peak at around 10 $M_\odot$ at the $2\sigma$ level, but it seems more likely that uncertainties larger than the Poisson errors are responsible.

An attempt to constrain the IMF (and SFR history) in five nearby irregular galaxies, using a combination of luminosity functions and color-magnitude diagrams has been given by Marconi et al. (1995 and references therein). A problem with this type of study is that, because it investigates relatively large regions (not individual clusters), the results depend on the inferred SFR history and on the adopted evolutionary models for massive stars. Besides, even though the LFs extend to intermediate masses, the main IMF results come from matching the LFs at large masses, where the degeneracy of colors with respect to effective temperatures should be a severe problem.

The resolution afforded by HST allows LFs to be estimated down to fairly low masses in Local Group galaxies, but a major problem is that the assumed star formation rate history strongly affects the derived IMF, similar to the problem for Milky Way field stars in the 1–1.5 $M_\odot$ range. An example is the recent HST study of stars in the Draco dwarf spheroidal galaxy by Grillmair et al. (1997), who estimate 95% Kolmogorov-Smirnov confidence levels for matching power law IMFs with the observed LF in the 0.6–0.9 $M_\odot$ range. The result for $\Gamma$ is –0.5 to –1.6, depending on the adopted age (10 to 15 Gyr). Considering other uncertainties (models, contamination, details of the star formation history) illustrates that it will be difficult to derive reliable IMFs from such data.

It appears that IMF estimates for even nearby galaxies beyond the Magellanic Clouds will have to rely on integrated light and chemical evolution constraints. A possible exception is the use of the statistics of the brightest and most massive stars in galaxies to (crudely) constrain the slope of the IMF (assuming it is a power law), an approach used by Schild & Maeder (1983) for nearby galaxies and by Hunter & Massey (1990) and Wilcots (1994) for H II regions in the Milky Way and LMC, respectively.

Figure 5. IMF index $\Gamma$ as a function of average log m for 61 clusters and associations taken from the papers discussed in the text. Filled and open symbols refer to the Milky Way and the LMC, respectively. In a few cases an individual cluster is shown as two points representing indices for different mass intervals.

## 7. Conclusions

The usual derivation of the field star IMF from the LF is of questionable use over several mass ranges. In the range 0.9–1.4 $M_\odot$ it is completely dominated by the assumed SFR history, and it makes no sense to speak of a field star IMF index. Above 15 $M_\odot$ and probably below about 0.1 $M_\odot$ the m-L relation is not single-valued, and comparison with evolutionary tracks in the H-R diagram is probably required. An additional problem for the very low-mass stars is that a significant fraction of them must be PMS stars, so for field stars the inferred IMF depends on the assumed SFR history, analogous to the problem in the 0.9–1.4 $M_\odot$ range. The corrections for stellar scale height and evolved stars are also large sources of uncertainty. Finally, the field stars sample objects out to different distances and maximum ages, depending on mass. For faint low-mass stars, for example, which can only be seen out to small distances, the young objects must have originated nearby, while the oldest objects may have originated kiloparsecs away. Thus if the IMF has variations in space or time, the field star IMF is a hopeless mixture which does not give a usefully defined "average IMF." For these reasons I believe that star clusters generally give a much better opportunity to study the IMF, and so I have spent most of this review discussing work related to clusters.

Keeping in mind the numerous sources of uncertainty persistently emphasized in this review, it is instructive to examine all the individual cluster results as a whole, and at face value. Fig. 5 shows the values of the IMF index $\Gamma$ for all the Milky Way (filled symbols) and LMC (open symbols) clusters discussed earlier as a function of the average value of log m for each case. Error bars are not shown for clarity, but typically range from 0.1 to 0.4 in $\Gamma$ with the larger uncertainties for the massive stars. It is important to remember that uncertainties associated with unresolved binaries and radial mass segregation should shift $\Gamma$ to smaller (steeper) values, and that the magnitude of both corrections will be larger for larger (flatter) $\Gamma$. The net effect, if important, would therefore be to compress the distribution of points to smaller $\Gamma$ with smaller dispersion.

The points in Fig. 5 represent only the papers discussed in this review. A large number of cluster studies (mostly earlier) were omitted for various reasons, including problems with the data, insufficient time to review so many papers in sufficient detail, and my lack of awareness of them. I also chose to omit any studies based on photographic photometry. A list of references on luminosity (or mass) functions for over 70 open clusters and associations (some of which overlap Fig. 5) can be found in Allen & Bastien (1995). However the power law IMF fits given in their Tables 2 and 4 include all the published data points, including those affected by incompleteness (at low masses) and evolution (near the turnoff mass), except for 7 clusters for which some low-mass points were omitted. (That is part of the reason why power laws were found to give such a poor fit.) Nevertheless, it would be worthwhile to carefully enlarge the dataset represented here in Fig. 5 and investigate correlations of IMF index with cluster properties (metallicity, age, star density, galactocentric radius,...) and compare with various theories both for the IMF itself and for the effects of cluster dynamical evolution, along the lines of Allen & Bastien (1995) and Kroupa (1995c,d). A consistent systematic study of the correction for unresolved binaries, applied to the specific cluster sample, is a prerequisite to such a project.

It is instructive to compare this plot with Fig. 34 of S86, which shows $\Gamma$ as a function of maximum mass for intermediate-mass stars in 75 clusters studied by Tarrab (1982). The reduction in scatter is perhaps encouraging, although it should be noted that the number of clusters in the same mass range is smaller for the sample shown in Fig. 5. Concentrating on the Milky Way data, the cluster of points at very small masses represents the Pleiades, Praesepe, $\rho$ Oph, NGC 2024, and L1495E. The result for the Pleiades is from Meusinger et al. (1996) and is consistent with the independent results of Williams et al. (1996) and Festin (1997). Given the large and numerous uncertainties, the agreement is surprisingly good, and suggests $\Gamma \approx 0$ to $-0.5$ for $m \approx 0.1$ to $0.8$ $M_\odot$, in agreement with most field star studies.

The two points at $m \sim 1$ are Praesepe and L1495E. At $m \approx 2$ the two low-$\Gamma$ points are the Pleiades and Be86, while the much larger $\Gamma \approx -0.1$ is for $\rho$ Oph. Above 3 $M_\odot$ the spread is still large, with $\Delta\Gamma \sim 1$. A similar spread is apparent for the LMC points. The very low point at $\Gamma = -3$ represents my hand fit to the LMC-SMC bridge result of Grondin et al. (1992) in the 1.4–2.5 $M_\odot$ range (see discussion in sec. 4).

Since the spread in $\Gamma$ at all masses above about 1 $M_\odot$ is so large, I see no basis for adopting some average value. In particular, I see no evidence for a clustering of points around the Salpeter value (–1.3), as has been claimed in several recent papers, except possibly for the LMC points between 3 and 9 $M_\odot$. Instead we are in the rather uncomfortable position of concluding that either the systematic uncertainties are so large that the IMF cannot yet be estimated, or that there are real and significant variations of the IMF index at all masses above about 1 $M_\odot$. If the latter interpretation is correct, then the IMF variations apparently do not depend much on metallicity, stellar density, or galactrocentric radius, but must either depend on some unknown combination of environmental variables or else the IMF, as usually conceived, is not a statistically well-defined probability function.

The last interpretation may not be as depressing as it sounds. Astronomers working on the IMF, like scientists in other fields, have always been rather desperate to arrive at a definite result, as evidenced by the trend in recent years to re-institute the Salpeter IMF despite the abundant evidence to the contrary in the literature reviewed here. Part of this urge, I think, is a result of a hope, or faith, that the complex and varied physical processes involved in star formation should somehow result in universal statistical properties. This belief is perhaps rooted in results from the velocity distribution found in kinetic theory and statistical mechanics and the power spectrum found in incompressible turbulence. However the IMF is by nature very different from these cases. There is no clear size scale separation for the star formation process, as is crucial for the universality of statistical mechanics, and the IMF is supposed to represent a one-point probability distribution, not a second order *moment* of a (two-point) probability distribution as is the case for the power spectrum of incompressible turbulence. In addition, the entire notion of ensemble average is problematic for spatial systems that lack statistical homogeneity (no structure on scales a significant fraction of the size of the region of interest) and isotropy. The fact that we are dealing with a highly inelastic system with no Galilean invariant quadratic conserved quantities should also be noted.

IMF theories have much to offer in this regard, if they would include a treatment of fluctuations rather than continuing to attempt to derive a mean IMF that fits whatever mean IMF seems to be prevalent in the literature.

Even rather simple models of the IMF may yield significant variations. Here is an example. In coalescence models, clouds or protostellar fragments are assumed to evolve through collisional interactions. The mass spectrum is derived by solving a kinetic equation that follows the collisional mass exchange between different mass intervals. A common result is a power law mass spectrum, although other solutions are possible. However it is well-recognized in the analogous problem of the size spectrum of terrestrial raindrops and aerosols that at each size or mass there is a probability distribution for the number of particles and the kinetic equation only gives the first moment (average) of this distribution. A large number of papers (see J. Atmos. Sci. over the past twenty years) have addressed this problem. Although it is usually not possible to derive this probability functional in closed form, it is certainly a problem accessible to simulations. Thus I propose that, if the evidence for variations presented here is real, there is a better chance for testing theories than in the case of a universal IMF. A study of the theoretical literature over the past thirty years shows how easy it is for theorists to produce a power law IMF of any desired index based on a large variety of physical or geometrical models. Instead, I suggest that theorists should predict the magnitude and nature of variations in IMF (say the variance as a function of mass). If a model predicts a "universal" IMF that cannot account for the level of variations suggested by Fig. 5, its viability is questionable.

The above discussion is of little consolation to researchers in galactic evolution, who cannot postpone their work until the proper definition of a fluctuating IMF is available. For this purpose I make the following suggestion. Perform every calculation using at least two disparate choices of IMF and do not assume that some "best IMF" is known. I think that if there is one conclusion that has emerged from the studies reviewed here, it is that the very low-mass IMF is relatively flat and stable in the solar neighborhood (a notable exception may be the Orion Nebula Cluster), and is steeper above about 1 $M_\odot$. Based mostly on the Milky Way work discussed here, the older cluster studies in S86, and the nature of effects due to unresolved binaries and radial mass segregation, my impression is that the IMF is steeper for intermediate-mass stars than for massive stars. If forced to choose an IMF for use in galactic evolution studies, I would suggest the following three-segment power law form (which is not meant to be taken as implying that the IMF is a power law over any mass range):

$$\Gamma = \begin{array}{ll} -0.2 \pm 0.3 & \text{for 0.1 to 1 } M_\odot \\ -1.7 \pm 0.5 & \text{for 1–10 } M_\odot \\ -1.3 \pm 0.5 & \text{for 10–100 } M_\odot \end{array}$$

The symbol "±" refers to the rough range of derived indices in each mass range, and may be interpreted as a measure of either the empirical uncertainties or the real IMF variations (or both). I am aware of no theory that would naturally produce an IMF that is steepest at intermediate masses, and leave this possibility as a challenge to theorists.

**Acknowledgments.** I thank Pavel Kroupa for informative and friendly correspondence that caused me to think more about the problems discussed in this paper, and Betty Friedrich for her assistance and patience in the preparation of this manuscript. This work was supported by NASA Grant NAG 5-3107.

# References

Alcalá, J. M., Krautter, J., Covino, E., Neuhäuser, R., Schmitt, J.H.M., & Wichmann, R. 1997, A&A, 319, 184

Allard, F., Hauschildt, P.H., Alexander, D.R., & Starrfield, S. 1997, ARAA, 35, 137

Allen, C. W. 1973, Astrophysical Quantities (London: Atlone Press)

Allen, E. J. & Bastien, P. 1995, ApJ, 452, 652

Alexander, D.R., Brocato, E., Cassisi, S., Castellani, V., Ciacio, F., & Degli'Innocenti, S. 1997, A&A, 317, 90

Aparicio, A., Alfaro, E. J., Delgado, A. J., Rodriguez-Ulloa, J. A., & Cabrera-Caño, J. 1993, AJ, 106, 4

Baraffee, I., Chabrier, G., Allard, F., & Hauschildt, P. H. 1995, ApJ, 446, L35

Banks, T., Dodd, R. J, & Sullivan, D. J. 1995, MNRAS, 274, 1225.

Barry, D. 1988, ApJ, 334, 436

Basu, S., & Rana, N. C. 1992, ApJ, 393, 373

Belikov, A. N. & Piskunov, A.E. 1997, ARep, 41, 28

Bencivenni, D, Brocato, E, Buananno, R., & Castellani, V. 1991, AJ, 102(1), 137

Bessel, M. S., & Stringfellow, G. S. 1993, ARAA, 31, 433

Brandl, B., Sams, B. J., Bertoldi, F. Eckart, A., Genzel, R, Drapatz, S., Hofmann, R., Lowew, M., & Quirrenbach, A. 1996, ApJ, 466, 254

Bresolin, F., Kennicutt, R.C., & Stetson, P.B. 1996, AJ, 112, 1009

Brown, A.G.A., de Geus, E. J., & de Zeeuw, P. T. 1994, A&A, 289, 101

Bryja, C. 1994, Ph.D. dissertation, U. Minnesota

Burrows, A., Hubbard, W. B., Saumon, D., & Lunine, J. I. 1993, ApJ, 406, 158

Carpenter, J. M., Meyer, M. R., Dougados, C., Strom, S. E., & Hillenbrand, L. A. 1997, AJ, 114, 198

Chabrier, G., Baraffee, I., & Plez, B. 1996, ApJ, 459, L91

Comeron, F., Rieke, G. H., & Rieke, M. J. 1996, ApJ, 473, 294

Comeron, F., Torra, J., & Rieke, G. H. 1996, A&A, 308, 565

De Marchi, G., & Paresce, F. 1997, ApJ, 476, L19

Deeg, H. J. & Ninkov, Z. 1996, A&AS, 119, 221

D'Antona, F. & Mazzetelli, I. 1983, A&A, 127, 149

D'Antona, F. & Mazzetelli, I. 1996, ApJ, 456, 329

Eaton, N., Adams, D. J., & Giles, A. B. 1984, MNRAS, 208, 241

Elson, R. A., Fall, M., & Freeman, K. C. 1989, Ap. J., 336, 73

Festin, L. 1997, A&A, 322, 455

Fletcher, A.B. & Stahler, S.W. 1994a, ApJ, 438, 313

Fletcher, A.B. & Stahler, S.W. 1994b, ApJ, 438, 329

de la Fuente Marcos, M. R. 1997, A&A, 322, 764

Garmany, C.D., Conti, P.S., & Chiosi, C. 1982, ApJ, 263,777

Gilmore, G. & Reid, N. 1983, MNRAS, 202, 1025

Giovannetti, P., Caux, E., Nadeau, D., & Monin, J.-L. 1997, A&A, in press

Gould, A., Bahcall, J. N., & Flynn, C. 1996, ApJ, 465, 759

Gould, A., Bahcall, J. N., & Flynn, C. 1997, ApJ, 482, 913

Grillmair, C. J., Mould, J. R., et al. 1997, preprint (astro-ph/9709259)

Grondin, L., Demers, S., & Kunkel, W. E. 1992, AJ, 103(4), 1234

Haywood, M. 1994, A&A, 282, 444

Haywood, M., Robin, A. C., & Creze, M. 1997a, A&A, 320, 428

Haywood, M., Robin, A. C., & Creze, M. 1997b, A&A, 320, 440

Henry, T. J. & McCarthy, D. W. 1993, AJ, 106, 773

Hill, R. J., Madore, B. F., & Freedman, W. L. 1994, ApJ, 429, 204

Hill, R. S., Cheng, K.-P., Bohlin, R. C., O'Connell, R. W., Roberts, M. S., Smith, A. M., & Stecher, T. P. 1995, ApJ, 446, 622

Hillenbrand, L. A. 1997, AJ, 113, 1733

Hunter, D. A. 1995, RevMexAA (Serie de Conferencias), 3, 1

Hunter, D. A., Baum, W. A., ONeil, E. J., Jr., & Lynds, R. 1996, ApJ, 456, 174

Hunter, D. A., Baum, W. A., ONeil, E. J., Jr., & Lynds, R. 1996, ApJ, 468, 633
Hunter, D. A., Light, R. M., Holtzman, J. A., Lynds, R., O'Neil, E. J. Jr., & Grillmair, C. J. 1997, ApJ, 478, 124
Hunter, D. A. & Massey, P. 1990, AJ, 99 (3), 846
Hunter, D. A., O'Neill, E. J., Lynds, R., Shaya, E. J., Groth, E. J., & Holtzman, J. A. 1996, ApJ, 459, L27
Hunter, D. A. & Plummer, J. D. 1996, ApJ, 462, 732
Hunter, D. A., Shaya, E. J., Holtzman, J. A., Light, R. M., O'Neil, E. J., & Lynds, R. 1995, ApJ, 448, 179
Jarrett, T. H., Dickman, R. L., & Herbst, W. 1994, ApJ, 424, 852
Jones, B. F. & Stauffer, J. R. 1991, AJ, 102, 1080
Kroupa, P. 1995a, ApJ, 453, 350
Kroupa, P. 1995b, ApJ, 453, 358
Kroupa, P. 1995c, MNRAS, 277, 1491
Kroupa, P. 1995d, MNRAS, 277, 1522
Kroupa, P. 1997, in Brown Dwarfs and Extrasolar Planets, ed. R. Rebolo, M.R. Zapatero Osorio, & E. Martin, in press.
Kroupa, P. & Tout, C. A. 1997, MNRAS, 287, 402
Kroupa, P., Tout, C. A., & Gilmore, G. 1990, MNRAS, 244, 76
Kroupa, P., Tout, C. A., & Gilmore, G. 1991, MNRAS, 251, 293
Kroupa, P., Tout, C. A., & Gilmore, G. 1993, MNRAS, 262, 545
Luhman, K. L. & Rieke, G. H. 1997, ApJ, in press
Maeder, A. & Conti, P. S. 1994, ARAA, 32, 227
Malkov, O.Y., Piskunov, A.E., & Shpil'kina, D.A. 1997, A&A, 320,79
Malumuth, E. M., Waller, W. H., & Parker, J. W. 1996, AJ, 111(3), 1128
Marconi, G., Tosi, M., Greggio, L., & Focardi, P. 1995, AJ, 109, 173
Massey, P. 1985, PASP, 97,5
Massey, P. 1997, in Stellar Astrophysics for the Local Group: A First Step to the Universe (Cambridge Univ. Press), in press
Massey, P., Johnson, K. E., & DeGioia-Eastwood, K 1995, ApJ, 454, 151
Massey, P., Armandroff, T. E., Pyke, R., Patel, K., & Wilson, C. D. 1995, AJ, 110, 2715
Massey, P., Lang, C. C., Degioia-Eastwood, K., & Garmany, C. D. 1995, ApJ, 438, 188
Massey, P., & Hunter, D. A. 1997, preprint
Mateo, M. 1988, ApJ, 331, 281
Megeath, S. T. 1996, A&A, 311, 135
Megeath, S. T., Herter, T., Beichman, C., Gautier, N., Hester, J. J., Rayner, J., & Shupe, D. 1996, A&A, 307, 775
Méra, D, Chabrier, G., & Baraffee, I. 1996, ApJ, 459, L87
Meusinger, H., Schilbach, E., & Souchay, J. 1996, A&A, 833, 844

Meyer, M. R. 1996, Ph.D. dissertation, U. of Massachusetts (see 1996, PASP, 108, 380)

Meusinger, M., Schilbach, E., & Souchay, J. 1996, A&A, 312, 833

Miller, G. E., & Scalo, J. M. 1979, ApJS, 41, 513

Ninkov, Z, Bretz, D. R., Easton, R. L. Jr., & Shure, M. 1995, AJ, 110, 5

Noh, H. & Scalo, J. 1990, ApJ, 352, 605

Oey, M. S., & Massey, P. 1995, ApJ, 452, 210

Oey, M. S. 1996, ApJ, 465, 231

Patat, F. & Carraro, G. 1995, MNRAS, 272, 507

Petitjean, P., Webb, J. K., Rauch, M., Carswell, R. F., & Lanzetta, K. 1993, MNRAS, 262, 499

Piskunov, A. E. & Belikov, A. N. 1996, AstL, 22, 466

Piskunov, A. E. & Malkov, O.Yu. 1991, A&A, 247, 87

Phelps, R.L. & Janes, K.A. 1993, AJ., 106, 1870

Rana, N. C. 1987, A&A, 184, 104

Reid, I. N. & Gizis, J. E. 1997, AJ, 113, 2246

Robin, A., & Crézé, M. 1986, A&A, 157, 71

Rocha-Pinto, H.-J, & Maciel, W. J. 1997, MNRAS, 289, 882

Sagar, R. & Richtler, T. 1991, A&A, 250, 324

Santiago, B. X., Gilmore, G., & Elson, R. A. W. 1996, MNRAS, 281, 871

Scalo, J. M. 1986, FCPh, 11, 1

Scalo, J. M. 1987, in Galaxy Evolution, ed. J. Palous (Astronomical Institute: Prague), 101

Scalo, J. M. 1987, in Starbursts and Galaxy Evolution, ed. T. Montmerle (Paris: Editions Frontiére), p. 445

Schild, H. & Maeder, A. 1983, A&A, 127, 238

Stobie, R. S., Ishida, K., & Peacock, J. A. 1989, MNRAS, 238, 709

Strom, K. M., Kepner, J., & Strom, S. E. 1995, AJ, 438, 813

Strom, R. M., & Strom, S. E. 1994, ApJ, 424, 237

Subramaniam, A. & Sagar, R. 1995, A&A, 297,695

Tarrab, I. 1982, A&A, 109, 285

Tinney, C. G., Mould, J. R., & Reid, I. N. 1992, ApJ, 396, 173

Twarog, B. A. 1980, ApJ, 242, 242

Vallenari, A., Bomans, D. J. & de Boer, K. S. 1993, A&A, 268, 137

Van Buren, D. 1985, ApJ, 294, 567

Vázquez, R. A., Baume, G., Feinstein, A., & Prado, P. 1997, A&AS, 124,13

Vesperini, E. & Heggie, D. C. 1997, MNRAS, 289, 898

Vio, R., Fasano, G., Lazzarin, M., & Lessi, O., 1994, A&A, 289, 640

von Hippel, T., Gilmore, G., Tanvir, N., Robinson, D., & Jones, D. 1996, AJ, 112,192
35

Wielen, R., Jahreiss, H., & Kruger, R. 1983, in IAU Colloq. 76, Nearby Stars and the Stellar Luminosity Function, ed. A.G.D. Philip & A. R. Upgren (Schenectady: L. Davis), p. 163

Wilcots, E. M. 1994, AJ, 108, 1674

Wilking, B. A. & Lada, C. J. 1983, AJ, 274, 698

Will, J.-M., Bomans, D. J., & de Boer, K. S. 1995a, A&A, 295, 54

Will, J.-M., Vázquez, R. A., Feinstein, A., & Seggewiss, W. 1995b, A&A, 301, 396

Will, J.-M., Bomans, D. J., & Dieball, A. 1997, A&ASupp, 123, 455

Williams, D. M., Comeron, F., Rieke, G. H., and Rieke, M. J. 1995, ApJ, 454, 144.

Williams, D. M., Boyle, R. P., Morgan, W. T., Rieke, G. H., Stauffer, J. R., & Rieke, M. J. 1996, ApJ, 464, 238

Williams, D. M., Rieke, G. H., & Stauffer, J. R. 1995, ApJ, 445, 359

Willmer, C.N.A. 1997, AJ, 114, 898

Zinnecker, H., McCaughrean, M.J., & Wilking, B.A. 1993, in Protostars and Planets III, ed. E.H. Levy & J.I. Lunine (Tucson: U. of Arizona Press), 429

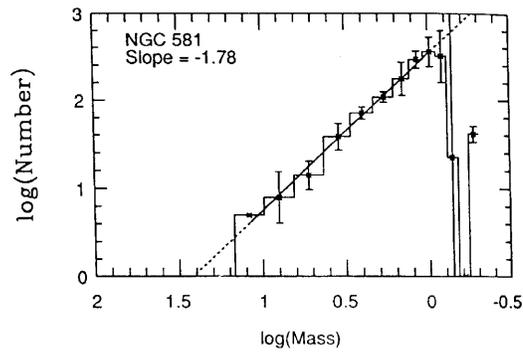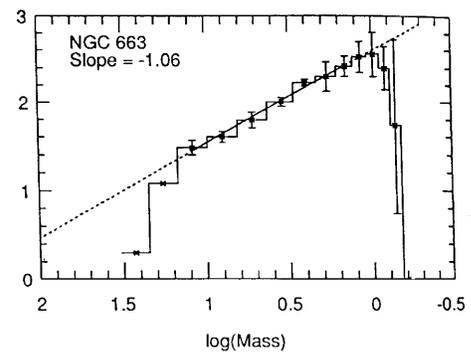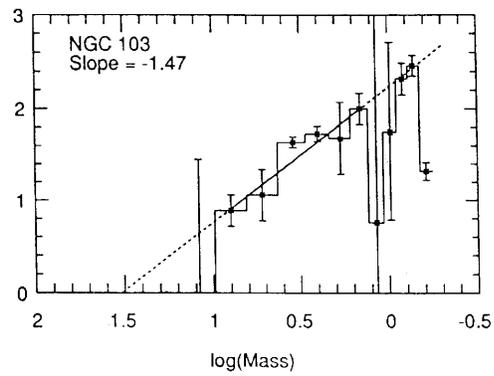

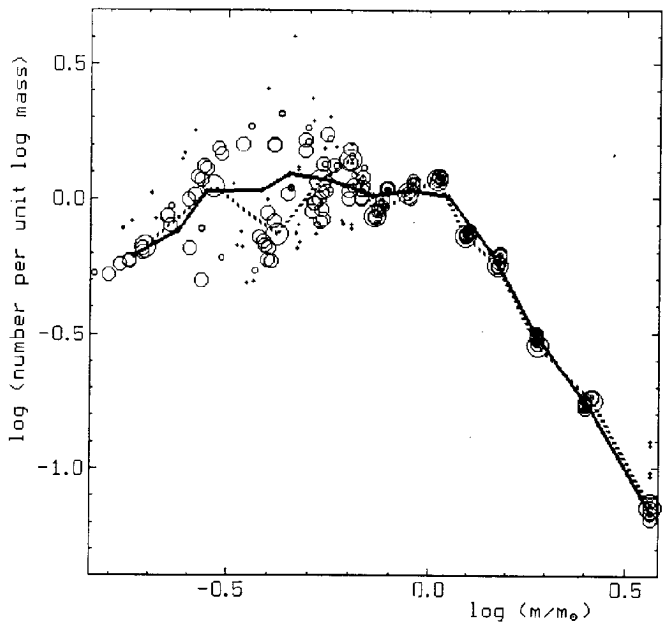

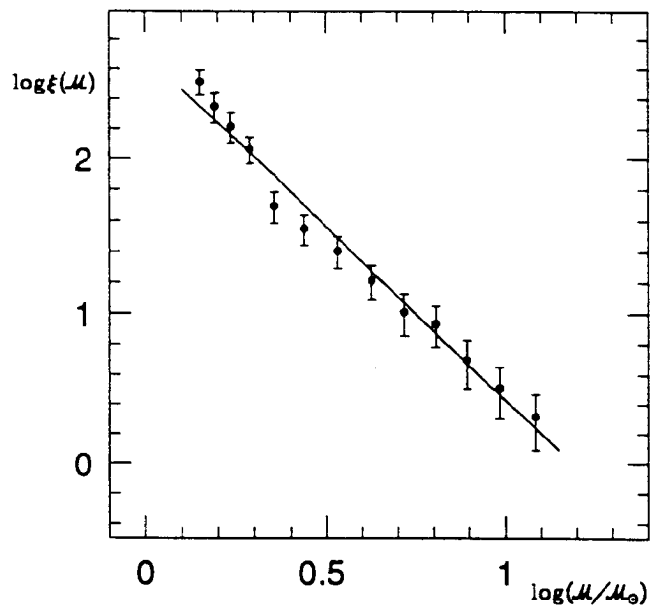

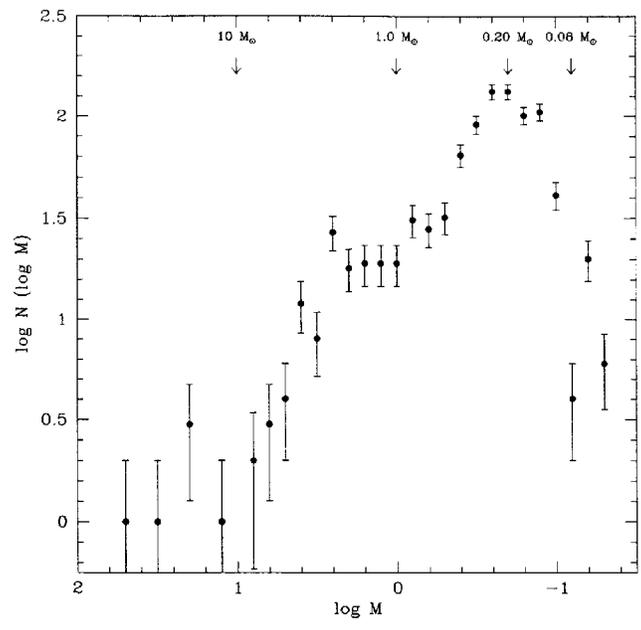

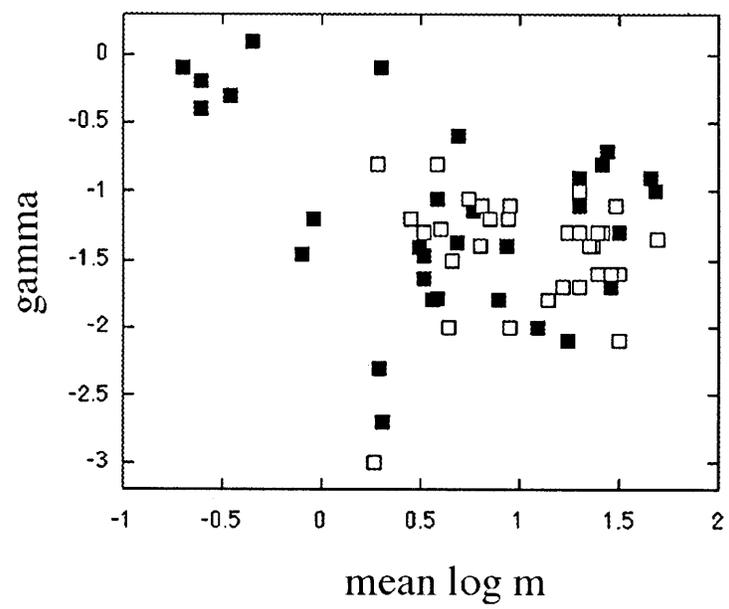